\documentclass[12pt]{article}
\usepackage{ametsoc}

\newcommand{\myabstract}{The new ZRP  wind input source term  \citep{R8} is checked for its consistency via numerical simulation of Hasselmann equation. The results are compared to 
field experimental data, collected at different sites around the world, and theoretical predictions of self-similarity analysis. Good agreement is obtained for limited fetch and time domain statements.}

\begin{document}

\title{\textbf{\large{On ZRP wind input term consistency in Hasselmann equation}}}

\author{
\textsc{Zakharov V.E.}\\
\centerline{\textit{\footnotesize{Department of Mathematics, University of Arizona, Tucson, AZ 85721, USA}}}\\
\centerline{\textit{\footnotesize{Novosibirsk State University, Novosibirsk, 630090, Russia}}}\\
\centerline{\textit{\footnotesize{Lebedev Physical Institute RAS, Leninsky 53, Moscow 119991, Russia}}}\\
\centerline{\textit{\footnotesize{Waves and Solitons LLC, 1719 W. Marlette Ave., Phoenix, AZ 85015, USA}}}\\
\textsc{Resio D.}\\
\centerline{\textit{\footnotesize{Taylor Engineering Research Institute, University of North Florida}}}\\
\centerline{\textit{\footnotesize{1 UNF Drive, Jacksonville, FL 32224, USA}}}\\
\textsc{Pushkarev A.} \thanks{Corresponding author, E-mail: \textit{dr.push@gmail.com}} \\
\centerline{\textit{\footnotesize{Waves and Solitons LLC, 1719 W. Marlette Ave., Phoenix, AZ 85015, USA}}}\\
\centerline{\textit{\footnotesize{Novosibirsk State University, Novosibirsk, 630090, Russia}}}\\
\centerline{\textit{\footnotesize{Lebedev Physical Institute RAS, Leninsky 53, Moscow 119991, Russia}}}
}

\ifthenelse{\boolean{dc}}
{
\twocolumn[
\begin{@twocolumnfalse}
\amstitle

% Start Abstract (Enter your Abstract above.  Do not enter any text here)
\begin{center}
\begin{minipage}{13.0cm}
\begin{abstract}
\myabstract
	\newline
	\begin{center}
		\rule{38mm}{0.2mm}
	\end{center}
\end{abstract}
\end{minipage}
\end{center}
\end{@twocolumnfalse}
]
}
{
\amstitle
\begin{abstract}
\myabstract
\end{abstract}
\newpage
}

%%%%%%%%%%%%%%%%%%%%%%%%%%%%%%%%%%%%%%%%%%%%%%%%%
					MAIN BODY OF PAPER
%%%%%%%%%%%%%%%%%%%%%%%%%%%%%%%%%%%%%%%%%%%%%%%%%

\section{Introduction}

The scientific description of wind driven wave seas, inspired by solid state physics statistical ideas (see, for instance, \cite{R3}), was proposed by \cite{R1,R2} in the form of kinetic equation for waves:
\begin{equation}
\label{HE}
\frac{\partial \varepsilon }{\partial t} +\frac{\partial \omega _{k} }{\partial \vec{k}} \frac{\partial \varepsilon }{\partial \vec{r}} =S_{nl} +S_{in} +S_{diss}
\end{equation}
where $\varepsilon =\varepsilon (\omega_k,\theta,\vec{r},t)$ is the wave energy spectrum, as a function of wave dispersion $\omega_k =\omega (k)$, angle $\theta$, two-dimensional real space coordinate $\vec{r}=(x,y)$ and time $t$. $S_{nl}$, $S_{in}$ and $S_{diss}$ are nonlinear, wind input and wave-breaking dissipation source terms, respectively. Hereafter, only the deep water case, $\omega=\sqrt{g k }$ is considered, where $g$ is the gravity acceleration and $k=|\vec{k}|$ is the absolute value of wavenumber $\vec{k}=(k_x,k_y)$. 

Since Hasselmann's work, Eq.({\ref{HE}}) has become the basis of operational wave forecasting models such as SWAN and Wavewatch III \citep{R9,R10}, While the physical oceanography community consents on the general applicability of Eq.({\ref{HE}}), there is no agreement on universal parameterizations of the source terms $S_{nl}$, $S_{in}$ and $S_{diss}$. 

The $S_{nl}$ term was derived by different methods from the primodary Euler equations for free surface incompressible potential flow of a liquid by  \cite{R1, R2} and \cite{R19}. It is complex nonlinear operator acting on $\varepsilon_k$, concealing hidden symmetries \citep{R4,R5}. \cite{R12} showed that those different forms are identical on the resonant surface
\begin{eqnarray}
\omega_{\vec{k_1}}+\omega_{\vec{k_2}} &=& \omega_{\vec{k_3}}+\omega_{\vec{k_4}} \\
\vec{k_1}+\vec{k_2} &=& \vec{k_3}+\vec{k_4}
\end{eqnarray}

The luxury of knowing the analytical expression for the $S_{nl}$ term, known in physical ocenography as XNL, is overshadowed by its computational complexity. Today, none of the operational wave forecasting models can afford performing XNL computations in real time. Instead, an approximation, known as DIA and its derivatives, are used \cite{}. The implication of such simplification is inclusion of a tuning coefficient in front of nonlinear term ; however, several publications have now shown that the form of the DIA does not provide a good approximation of the actual form of XNL \cite{}. Consequently other source terms must be adjusted to allow the model Eq.(\ref{HE}) to obtain desirable results.

In contrast to $S_{nl}$, the knowledge of the $S_{in}$ and $S_{diss}$ source terms is poor, so these now have many heuristic factors and terms included within them.  

The creation of a reliable, well justified theory of $S_{in}$ has been hindered by strong turbulence presence in the air boundary layer over the sea surface. Even one of the most crucial elements of this theory, the vertical distribution of horizontal wind velocity in the region closest to the ocean surface, where wave motions strongly interact with atmospheric motions, is still the subject of debate. The history the development of different $S_{in}$ wind input forms is full of misconceptions, such as: the "fractional derivative method",  wind driven waves studies in shallow basins, which fundamentally restrict the surface wave speed propagation, questionable data interpretations, and others.  As a result, the values of different wind input terms scatter by the factor of $300 \div 500 \%$ \citep{R11,R7}. Additional information on this detailed analysis of current state of wind input terms can be found   in \cite{R7}.  

Similar to the wind input term, there is little consent on the parameterization of the source dissipation term $S_{diss}$. The physical dissipation mechanism,  which most physical oceanographers agree on, is the effect of wave energy loss due to wave breaking, while there are also other dubious ad-hoc "long wave" dissipation source terms, having heuristically justified physical explanations. Wave breaking dissipation, known also as "white-capping dissipation", is an important physical phenomenon, not properly studied yet for the reasons of mathematical and technological complexity. Currently, there is not even agreement on the localization of wave breaking events in Fourier space. The approach currently utilized in operational wave forecasting models mostly relies on the dissipation, localized in the vicinity of the spectral energy peak. Recent numerical experiments show \citep{R21,R22}, however, that such approach does not pass most of the tests associated with the essentially nonlinear nature of the Eq.(\ref{HE}). 

Theoretical and experimental study of wave-breaking is extremely difficult task due to mathematical and technological complexities. \cite{R29,R30} achieved important results, but didn't accomplished the theory completely. \cite{R31} studied wave-breaking of the short waves, "squeezed" by surface currents, caused by longer waves, and showed that they become steep and unstable. Our explanation is simpler, but has the same consequiences:  the "wedge" formation, preceeding the wave breaking, causes the "fat tail" appearance is Fourier space. Subsequent smoothing of the tip of the wedge is equivalent to the "chopping off" of the developped high-frequency tail -- the sort of natural low-pass filtering -- leading to the loss of the wave energy. This effect is referred as "cigar cutting effect" \citep{R7}. Both scenarios have the same consequences of wave surface smoothing, and are indirectly confirmed by presented numerical experiments.

Instead of following the previous path of time-consuming numerical and field experiments, the authors of the current manuscript are pursuing an alternative approach to definition of $S_{in}$ and $S_{diss}$ terms . Based on the leading nonlinearity role in Eq.(\ref{HE}) (\citep{R14,R15}), it was desided to analyze a multi-parametric family of self-similar solutions of Eq.(\ref{HE}). The comparison with the results of field observations allowed to find a new wind input form, herein termed the ZRP $S_{in}$ wind input source term \citep{R8}. The wave breaking dissipation term $S_{diss}$ has been chosen in the form of "implicit dissipation" via Phillips $\sim f^{-5}$ spectral continuation tail. 

That framework reproduced the observations of a dozen of field experiments , i.e. self-similar exponents $p=1$ and $q=-0.3$ of power dependencies of total wave energy $\sim \chi^{p}$ and spectral peak frequency $\sim \omega^{q}$ along the fetch \citep{R7}.

In the following section we are describing the details of ZRP model and pertaining numerical results in time domain and duration limited statements. 

\section{Experimental evidence}

Here we examine the empirical evidence from around the world, which has been utilized to quantify energy levels within the equilibrium spectral range by \cite{R27}.  For convenience, we shall also use the same notation used by \cite{R27} in their study, for the angular averaged spectral energy densities in frequency and wavenumber spaces:
\begin{eqnarray}
\label{ZKspec2}
E_4(f) &=& \frac{2\pi \alpha_4 V g}{(2 \pi f)^4} \\
\label{ZKspec3}
F_4(k) &=& \beta k^{-5/2} 
\end{eqnarray}
where $f=\frac{\omega}{2 \pi}$, $\alpha_4$ is the constant, $V$ is some characteristic velocity and $\beta=\frac{1}{2}\alpha_4 V g^{-1/2}$. These notations are based on relation of spectral densities $E(f)$ and $F(k)$ in frequency $f=\frac{\omega}{2\pi}$ and wave-number $k$ bases:
\begin{eqnarray}
\label{EnRel}
F(k)=\frac{c_g}{2\pi}E(f)
\end{eqnarray}
where $c_g = \frac{d \omega}{d k} = \frac{1}{2 \cdot 2 \pi} \frac{g}{f}$ is the group velocity.

\begin{figure}[t]
\noindent\center\includegraphics[scale=1.4]{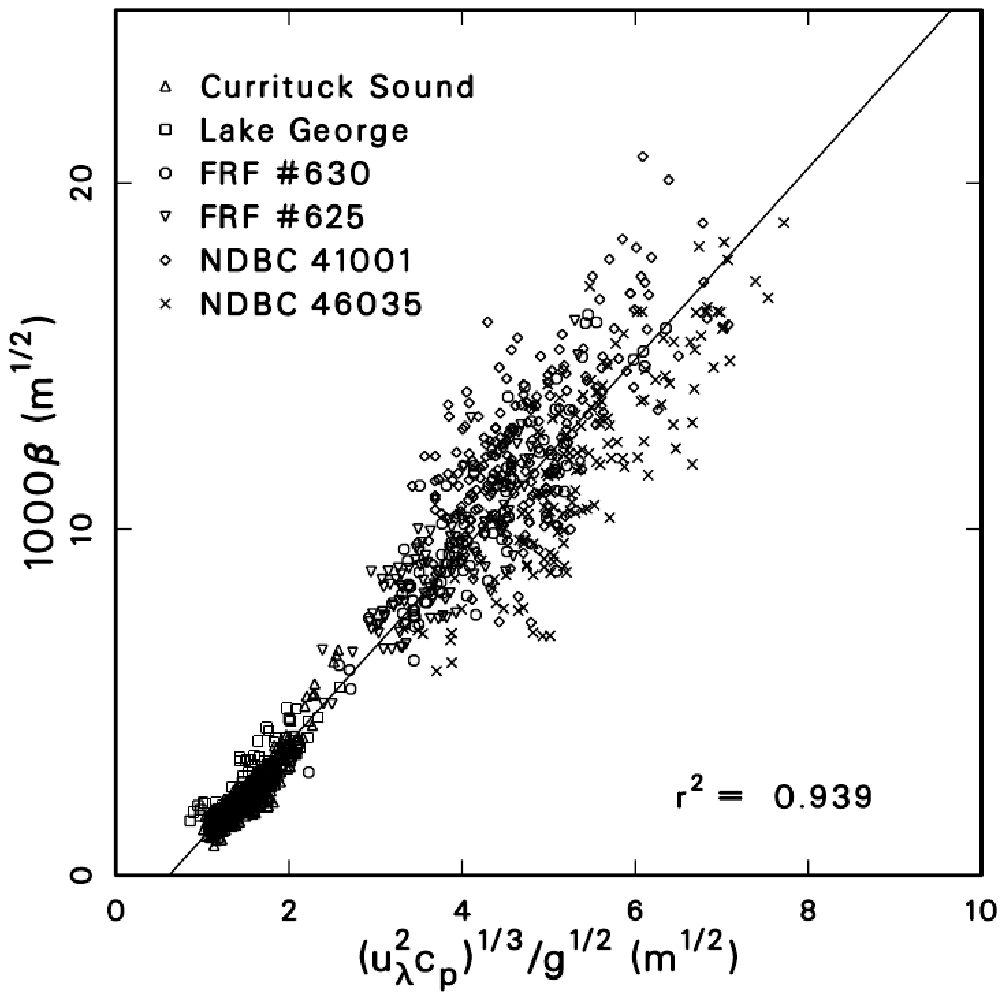} 
\caption{ Correlation of equilibrium range coefficient $\beta$ with $(u_\lambda^2 c_p)^{1/3}/g^{1/2}$ based on data from six disparate sources. Adapted from \cite{R27}} \label{RegLine}
\end{figure}

The notations in Eqs.(\ref{ZKspec2})-(\ref{ZKspec3}) are connected with the spectral energy density $\epsilon(\omega,\theta)$  through
\begin{eqnarray}
E(f) = 2 \pi \int \epsilon(\omega,\theta) d\theta
\end{eqnarray}
The \cite{R27} analysis showed that experimental energy spectra $F(k)$, estimated through averaging $<k^{5/2}F(k)>$, can be approximated by linear regression line as a function of $(u^2_\lambda c_p)^{1/3} g^{-1/2}$. Fig.\ref{RegLine} shows that the regression line 
\begin{eqnarray}
\beta=\frac{1}{2} \alpha_4 \left[ (u_\lambda^2 c_p)^{1/3}-u_0\right]g^{-1/2}
\end{eqnarray}
indeed, seems to be a reasonable approximation of these observations.

Here $\alpha_4=0.00553$, $u_0=1.93$ $m/sec$, $c_p$ is the spectral peak phase speed and $u_\lambda$ is the wind speed at the elevation equal to a  fixed fraction  $\lambda=0.065$ of the spectral peak wavelength $2\pi/k_p$, where $k_p$ is the spectral peak wave number. It is important to emphasize that \cite{R27} experiments show that parameter $\beta$ increases with development of the wind-driven sea, when $f_p$ decreases and $C_p$ increases. This observation is consistent with the weak turbulent theory, where $\beta ~ P^{1/3}$ \citep{R5} ($P$ is the wave energy flux toward small scales).

 \cite{R27} assumed that the near surface boundary layer can be treated as neutral and thus follows a conventional logarithmic profile   
\begin{eqnarray}
\label{LogProfile}
u_\lambda=\frac{u_\star}{\kappa} \ln \frac{z}{z_0}
\end{eqnarray}
having  Von Karman coefficient $\kappa=0.41$, where $z=\lambda \cdot 2\pi/k_p$ is the elevation equal to a fixed fraction $\lambda = 0.065$ of the spectral peak wavelength $2\pi/k_p$,  where $k_p$ is the spectral peak wave number, and $z_0=\alpha_C u_{\star}^2 /g$ subject to \cite{R26} surface roughness with $\alpha_C = 0.015$.

\section{Theoretical considerations}

Self-similar solutions of conservative kinetic equation 
\begin{eqnarray}
\label{ConsHasEq}
\frac{\partial \epsilon(\omega,\theta)}{\partial t} = S_{nl} 
\end{eqnarray}
were studied in \cite{R17}, \cite{R11}. In this chapter we study self-similar solutions of the forced kinetic equation
\begin{eqnarray}
\label{ForcedHasEq}
\frac{\partial \epsilon(\omega,\theta)}{\partial t} = S_{nl}+\gamma(\omega,\theta) \epsilon(\omega,\theta)
\end{eqnarray}
where $\epsilon(\omega,\theta)=\frac{2\omega^4}{g}N({\bf k},\theta)$ is the energy spectrum. For our purposes, it is sufficient to simply use the dimensional estimate for $S_{nl}$,
\begin{eqnarray}
S_{nl} \simeq \omega \left( \frac{\omega^5 \epsilon}{g^2} \right)^2 \epsilon
\end{eqnarray}
Eq.(\ref{ForcedHasEq}) has a self-similar solution if
\begin{eqnarray}
\gamma(\omega,\theta) = \alpha \omega^{1+s} f(\theta)
\end{eqnarray}
where $s$ is a constant. Looking for self-similar solution in the form 
\begin{eqnarray}
\epsilon(\omega,t) = t^{p+q} F(\omega t^q)
\end{eqnarray}
we find
\begin{eqnarray}
q&=&\frac{1}{s+1} \\
p&=&\frac{9 q-1}{2} = \frac{8-s}{2 (s+1)}
\end{eqnarray}
The function $F(\xi)$ has the maximum at $\xi\sim\xi_p$, thus the frequency of the spectral peak is 
\begin{equation}
\omega_p\simeq\xi_p t^{-q}
\end{equation}
The phase velocity at the spectral peak is
\begin{equation}
c_p = \frac{g}{\omega_p} = \frac{g}{\xi_p} t^q = \frac{g}{\xi_p} t^{\frac{1}{s+1}}
\end{equation}
According to experimental data, the main energy input into the spectrum occurs in the vicinity of the spectral peak, i.e. at $\omega\simeq \omega_p$. For $\omega>>\omega_p$, the spectrum is described by Zakharov-Filonenko tail
\begin{equation}
\epsilon(\omega) \sim P^{1/3} \omega^{-4}
\end{equation}
Here 
\begin{equation}
P = \int_0^\infty \int_0^{2 \pi} \gamma(\omega,\theta) \epsilon(\omega,\theta) d\theta
\end{equation}
This integral converges if $s<2$. For large $\omega$ 
\begin{eqnarray}
\epsilon(\omega,t) \simeq \frac{t^{p-3q}}{\omega^4} \simeq \frac{t^{\frac{2-s}{2(s+1)}}}{\omega^4}
\end{eqnarray}
More accurately
\begin{eqnarray}
\label{ZKspec1} 
\epsilon(\omega,t) &\simeq& \frac{\mu g}{\omega^4} u^{1-\eta} c_p^\eta g(\theta) \\
\eta &=& \frac{2-s}{2}
\end{eqnarray}
Now supposing $s=4/3$ and $\gamma\simeq \omega^{7/3}$, we get $\eta=1/3$ , which is exactly experimental regression line prediction. Because it is known from regression line on Fig.\ref{RegLine} that $\xi=1/3$, we immediately get $s=4/3$ and the wind input term
\begin{eqnarray}
\label{NewWindInput}
S_{wind} \simeq \omega^{7/3} \epsilon
\end{eqnarray}

At the end of the section, we present the summary  of important relationships.

Wave action $N$, energy $E$ and momentum $M$ in frequency-angle presentation are:
\begin{eqnarray}
N &=& \frac{2}{g^2} \int\omega^3 n d\omega d\phi \\
E &=& \frac{2}{g^2} \int\omega^4 n d\omega d\phi \\
M &=& \frac{2}{g^3} \int\omega^5 n \cos{\phi} d\omega d\phi
\end{eqnarray}

The self-similar relations  for duration limited case:

\begin{eqnarray}
\epsilon &=& t^{p+q} F(\omega t^q)	\\
\label{MagicRelationTime}
9q-2p &=& 1,\,\,\,\,p=10/7,\,\,\,\,q=3/7\,\,\,\,s=4/3	\\
N & \sim & t^{p+q}  \\
\label{EnTimeSelfSim}
E & \sim & t^p	\\
M & \sim & t^{p-q}	\\
\label{FrTimeSelfSim}
<\omega> &\sim& t^{-q}
\end{eqnarray}

The same sort  of self-similar analysis gives self-similar relations for fetch limited case: 

\begin{eqnarray}
\epsilon &=& \chi^{p+q} F(\omega \chi^q)	\\
\label{MagicRelationFetch}
10q-2p &=& 1,\,\,\,\,p=1,\,\,\,\,q = 3/10 \,\,\,\,s=4/3	\\
N & \sim & \chi^{p+q} 	\\
\label{EnFetchSelfSim}
E & \sim & \chi^p		\\
M & \sim & \chi^{p-q}	\\
\label{FrFetchSelfSim}
<\omega> &\sim& \chi^{-q}
\end{eqnarray}

\section{Numerical simulation}

To check the self-similar conjecture Eq.(\ref{NewWindInput}), we performed the series of numerical simulations of Eq.(\ref{HE}) in the spatially homogeneous time domain $\frac{\partial N}{\partial {\vec{r}}}=0$ and spatially inhomogeneous fetch limited  $\frac{\partial N}{\partial t} =0$ situations. The same wind input term Eq.(\ref{NewWindInput}) has been used in both cases in the form
\noindent
\begin{eqnarray}
&&S_{in}(\omega,\phi) = \gamma(\omega,\phi)\cdot \varepsilon(\omega,\phi) \\
&&\gamma = 0.05 \frac{\rho _{air} }{\rho _{water} } \omega \left(\frac{\omega }{\omega _{0} } \right)^{4/3} f(\theta ) \label{ZRP1} \\
&&f(\theta ) = \left\{\begin{array}{l} {\cos ^{2} \theta {\rm \; \; for\; -}\pi {\rm /2}\le \theta \le \pi {\rm /2}} \\ {0{\rm \; \; otherwise}} \end{array}\right. \\
&&\omega _{0} = \frac{g}{u_{10} }, \,\,\,  \frac{\rho _{air} }{\rho _{water} } =1.3\cdot 10^{-3}
\end{eqnarray} 
\noindent
where $u_{10}$ is the wind speed, $\rho_{air}$ and $\rho_{water}$ are the air and water density correspondingly. It is conceivable to use more sophisticated expression for $f(\theta)$, for instance $f(\theta) = q(\theta)-q(0)$. The wind speed $u_{10}$ is taken here as the speed at a reference level of 10 meters. To make comparison with experimental results of \cite{R27}, we used relation $u_{\star} \simeq u_{10}/28$ (see \cite{R28}) in Eq.(\ref{LogProfile}).

Both situations also need the knowledge of the dissipation term $S_{diss}$, which is taken into account as was proposed by \cite{R27}, where white-capping was introduced implicitly through an $f^{-5}$ energy spectral tail stretching in frequency range from $f_d =1.1$ to $f_{max} = 2.0$. To date, this approach has been confirmed by both experimental observations and numerical experiments to provide an effective direct cascade energy sink at high frequencies.

The nonlinear $S_{nl}$ term has been calculated in the exact XNL form.

\subsection{Time domain simulation}

All time domain numerical simulations shown here have been started from uniform noise energy distribution in Fourier space. 

\begin{figure}[t]
\includegraphics[scale=0.7]{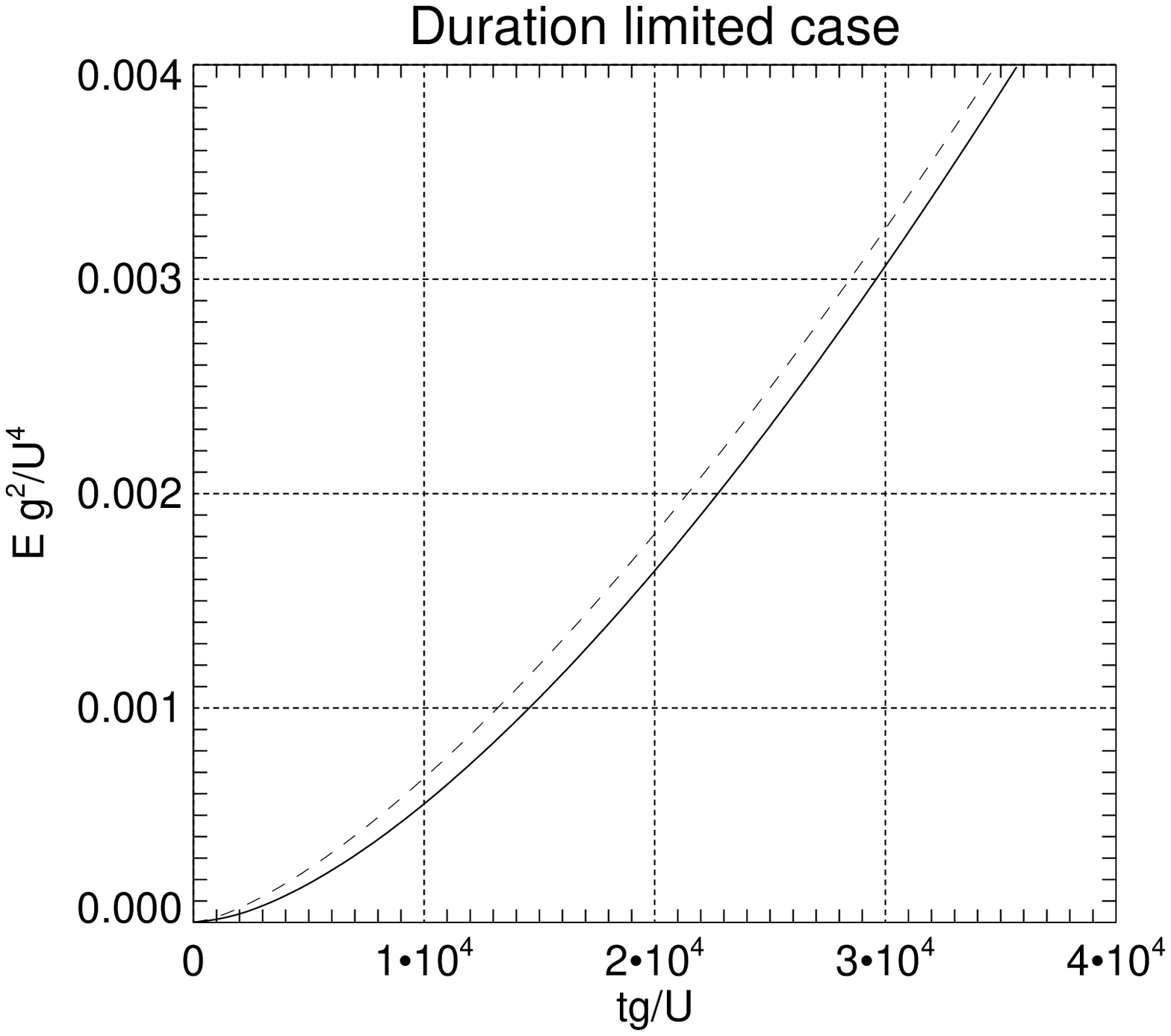}
\caption{
Dimensionless energy $Eg^2/U^4$ versus dimensionless time $tg/U$ for duration limited case. Dashed fit $1.3 \cdot 10^{-9} \left(tg/U\right)^{10/7}$
}\label{EnergyDimDL}
\end{figure}

\begin{figure}[t]
\includegraphics[scale=0.7]{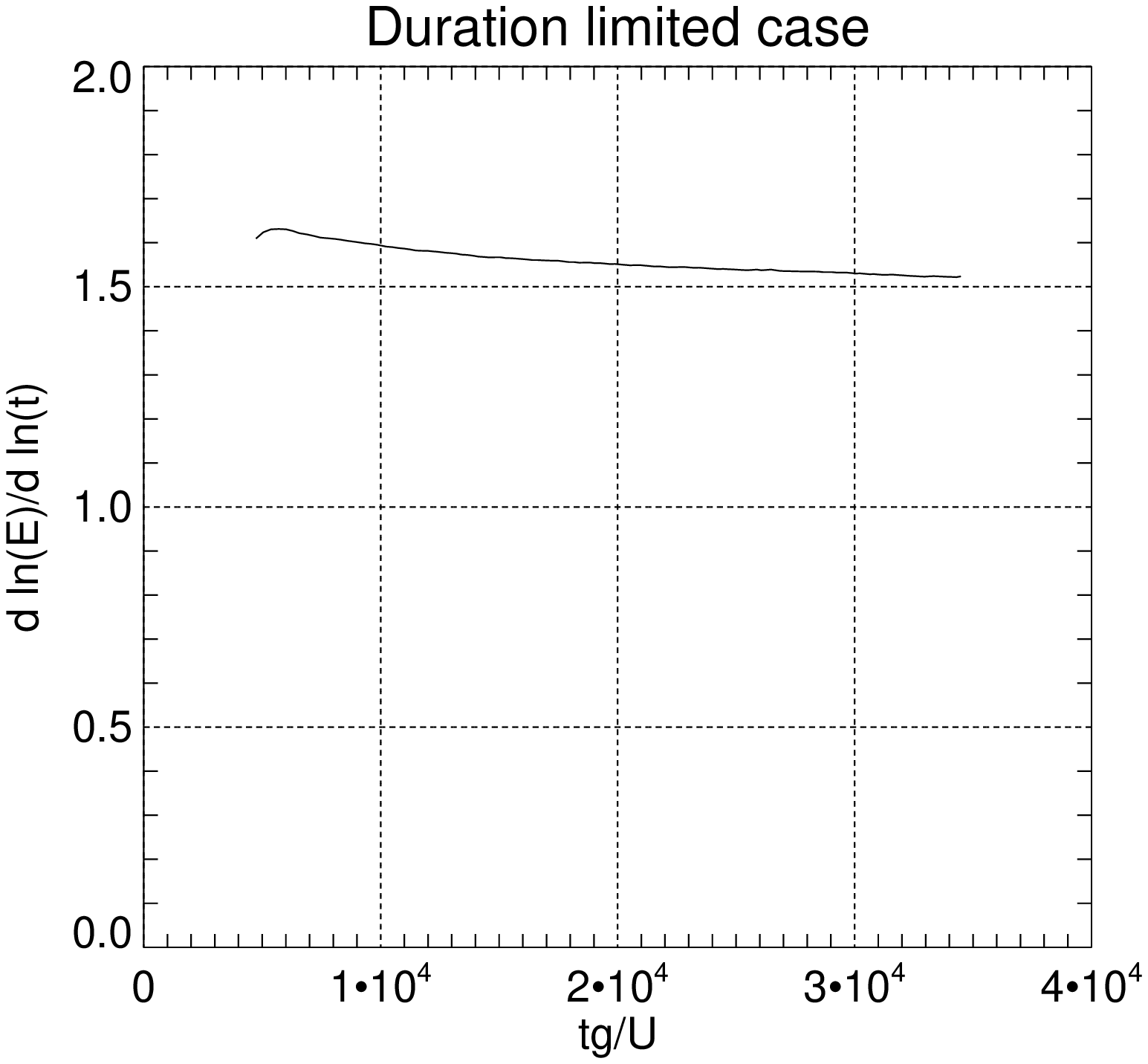}
\caption{
Local energy exponent power $p=\frac{d\ln{E} }{d\ln{t} }$ as a function of dimensionless time $tg/U$ for duration limited case.
}\label{EnergyIndexDL}
\end{figure}

Fig.\ref{EnergyDimDL} shows the total energy growth as a function of time close to self-similar prediction Eq.(\ref{EnTimeSelfSim}) for $p=10/7$, see Fig.\ref{EnergyIndexDL}.

\begin{figure}[t]
\includegraphics[scale=0.7]{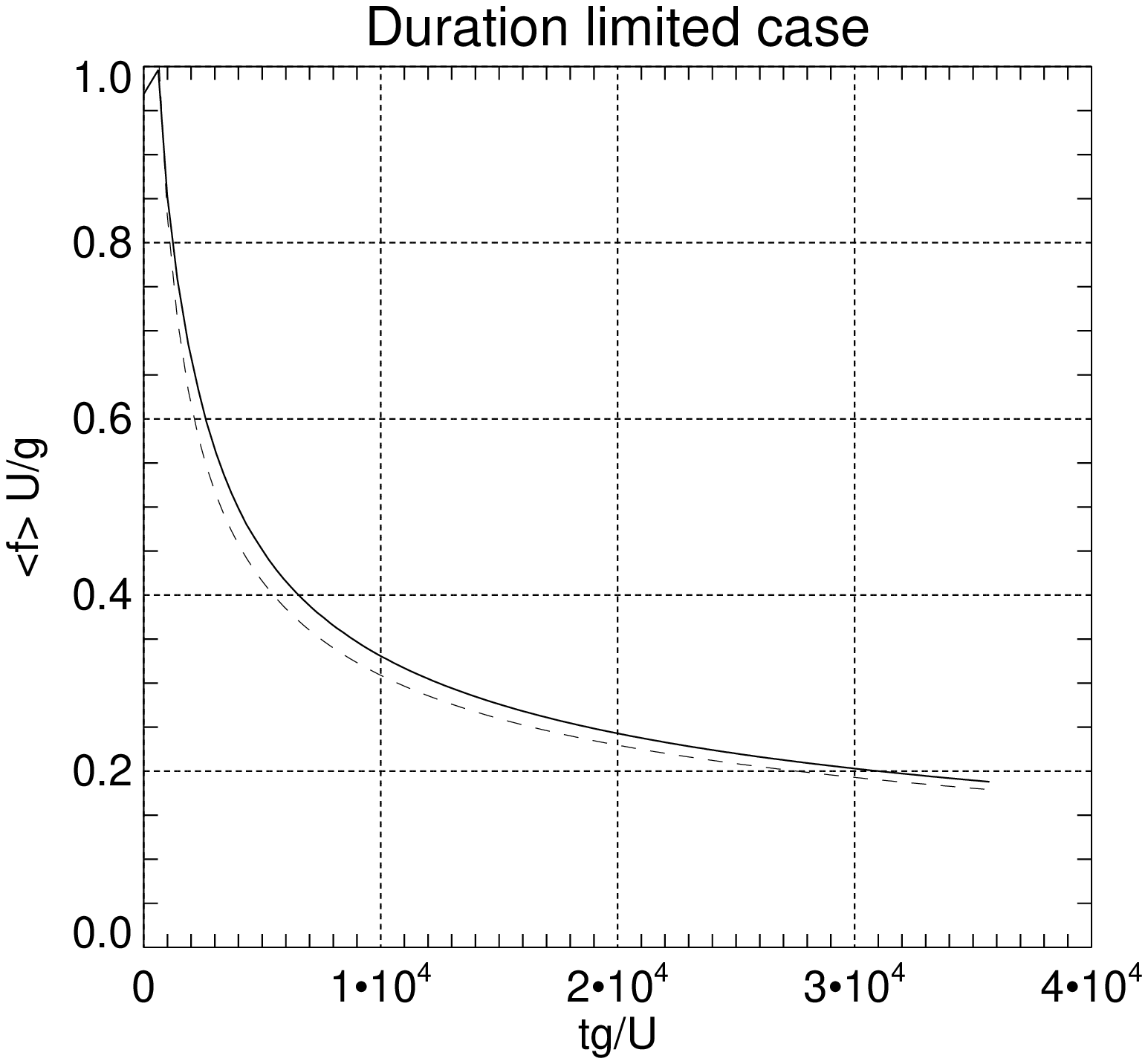}
\caption{
Dimensionless frequency $<f> \cdot U/g = E/N \cdot U/g$ versus dimensionless time $tg/U$ for duration limited case, dashed fit by $ 16.0 \cdot (tg/U)^{-3/7} $.
}\label{MeanFreqDimDL}
\end{figure}

\begin{figure}[t]
\includegraphics[scale=0.7]{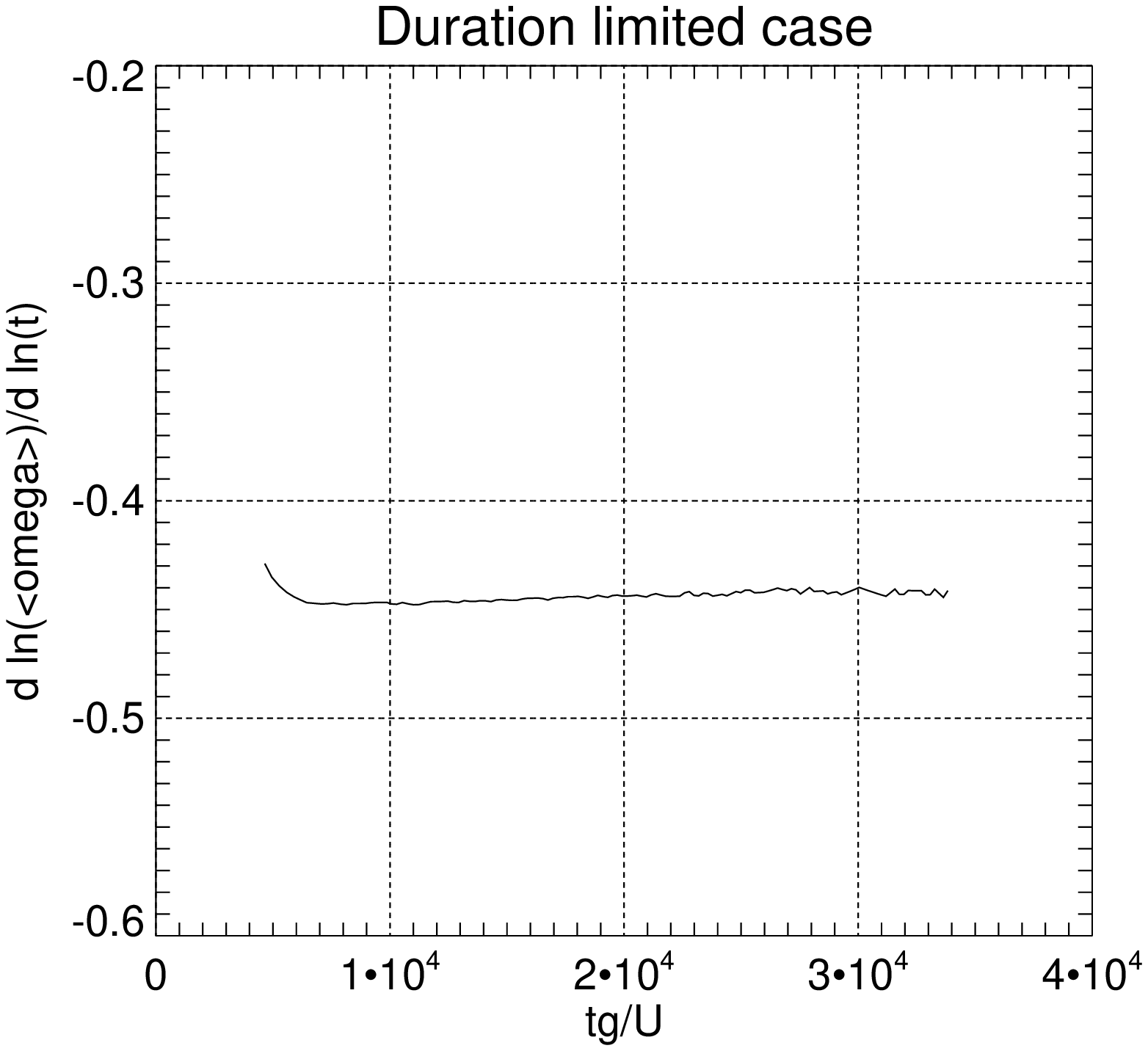}
\caption{
Local mean frequency exponent power $q=\frac{d\ln{<\omega>} }{d\ln{t} }$ as a function of dimensionless time $tg/U$ for duration limited case.
}\label{MeanFreqIndexDL}
\end{figure}

The dependence of mean frequency, on the fetch, shown in Fig.\ref{MeanFreqDimDL}  demonstrates  good correspondence with self-similar dependence Eq.(\ref{FrTimeSelfSim}), for $q = 3/7$, see Fig.\ref{MeanFreqIndexDL}.

\begin{figure}[t]
\includegraphics[scale=0.7]{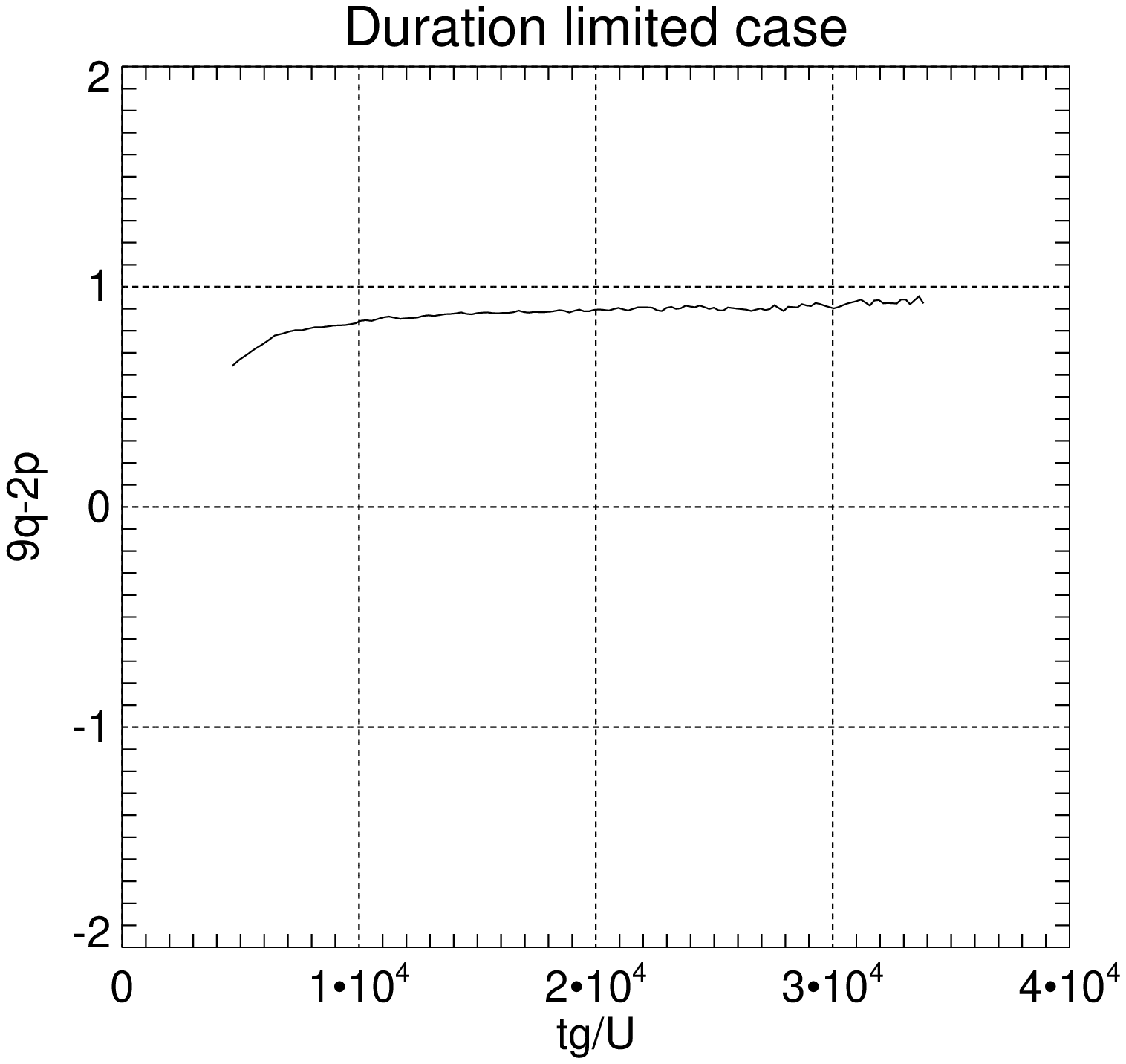}
\caption{
"Magic number" $9q-2p$ as a function of dimensionless time $tg/U$ for duration limited case.
}\label{MagicNumberDL}
\end{figure}

The check of consistency with the “magic number” $(9q-2p)$  \citep{R7}, is presented on Fig.\ref{MagicNumberDL}. This also agrees with the self-similar prediction Eq.{\ref{MagicRelationTime}}.

\begin{figure}[t]
\includegraphics[scale=0.7]{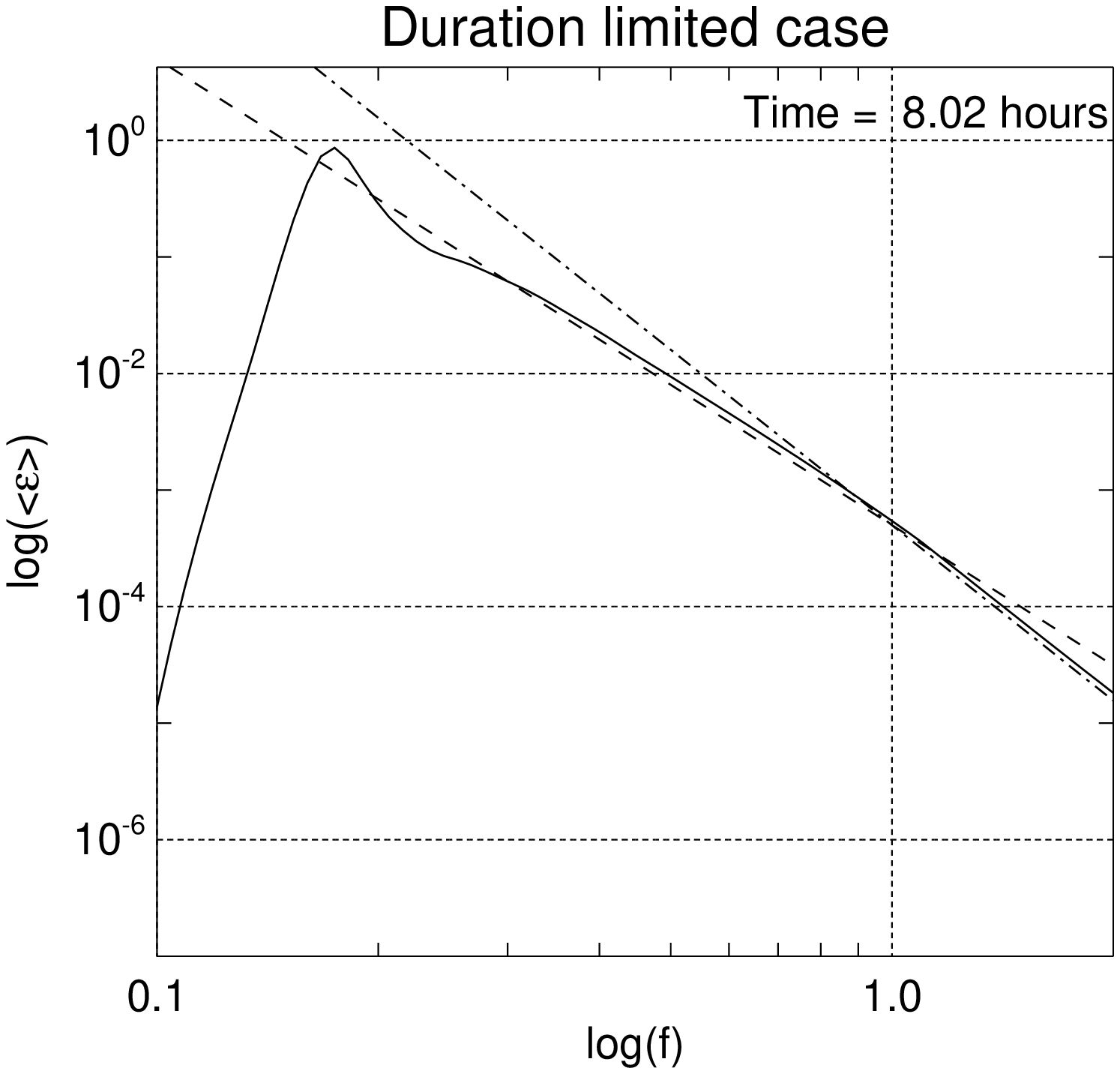}
\caption{
Decimal logarithm of the angle averaged spectrum as a function of the decimal logarithm of the frequency for duration limited case.
}\label{LogLog8hDL}
\end{figure}

Fig.\ref{LogLog8hDL} presents angle-integrated energy spectrum, as the function of frequency, in logarithmic coordinates. One can see that it consists of the segments of:

\begin{itemize}
\item{} spectral maximum area
\item{} spectrum $~\omega^{-4}$
\item{} Phillips high frequency tail $\omega^{-5}$
\end{itemize}

\begin{figure}[ht]
\includegraphics[scale=0.7]{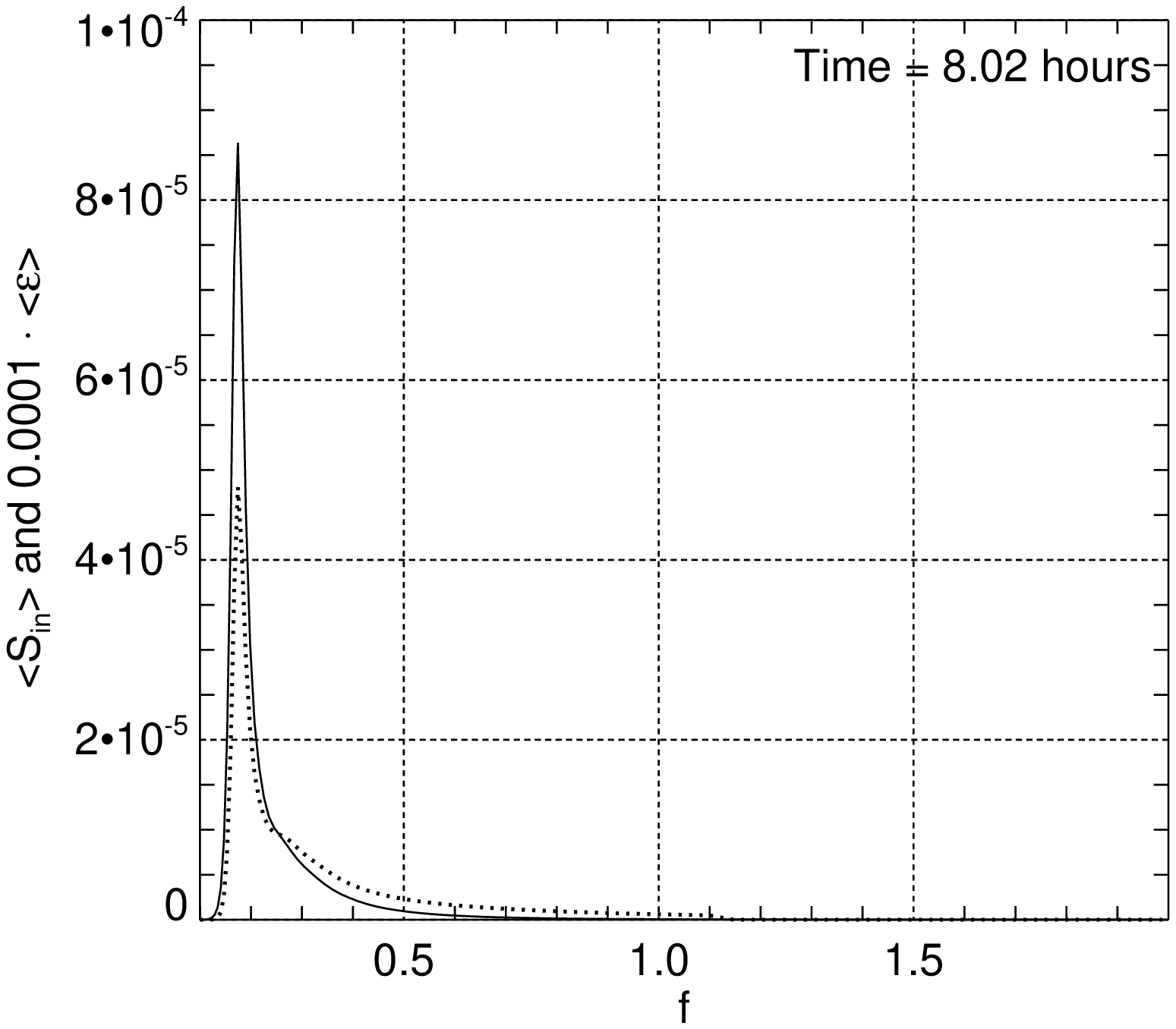}
\caption{
Typical, angle averaged, wind input function density $<S_{in}> = \frac{1}{2 \pi}\int \gamma(\omega,\theta) \varepsilon(\omega,\theta) d\theta$ and angle averaged spectrum $<\varepsilon> = \frac{1}{2\pi} \int \varepsilon(\omega,\theta) d\theta$ (solid line) as the functions of the frequency $f=\frac{\omega }{2\pi}$.
}\label{EnergyInputDurationLimited}
\end{figure}

\begin{figure}[ht]
\includegraphics[scale=0.7]{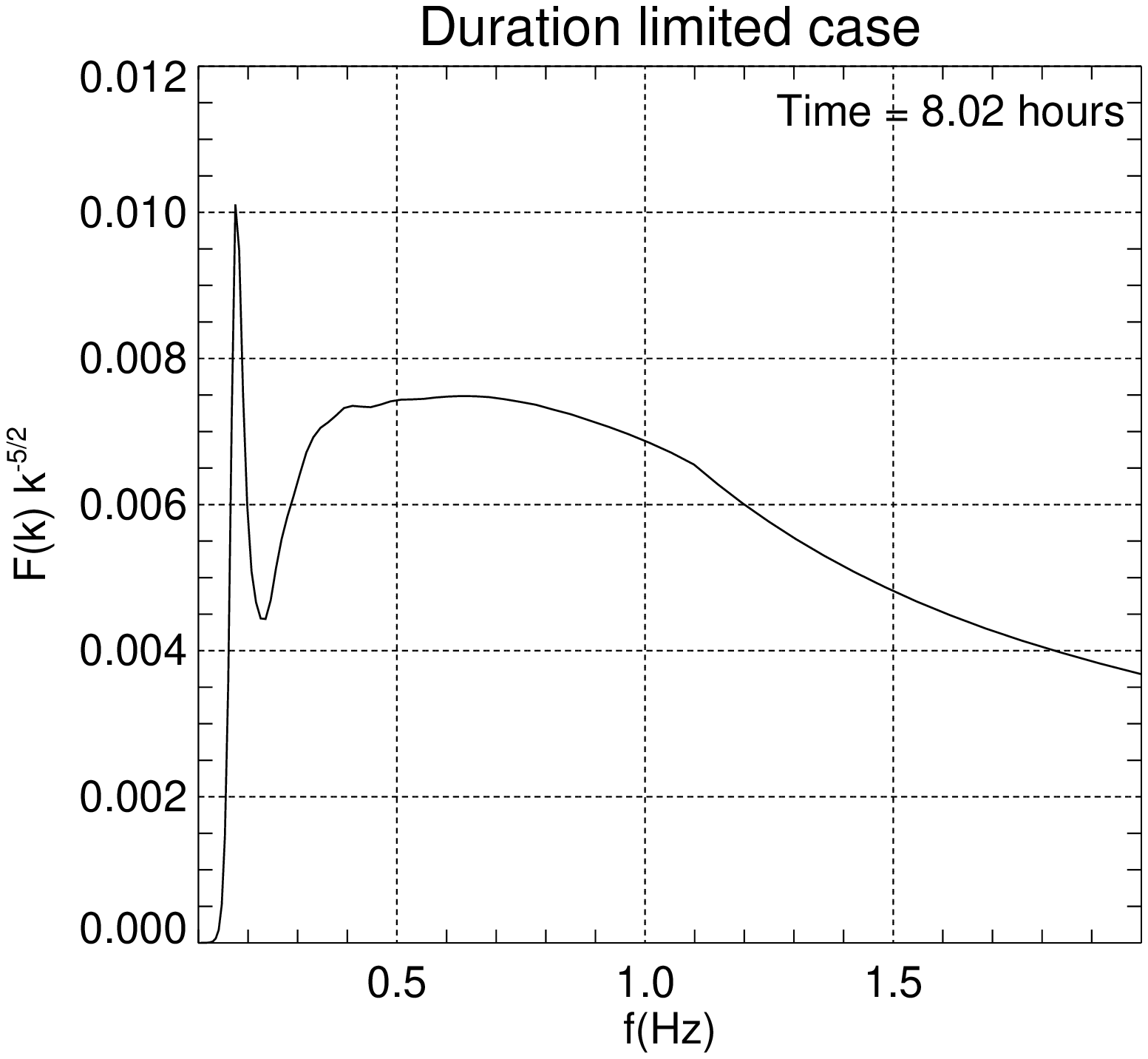}
\caption{
Compensated spectrum for duration limited case as a function of linear frequency $f$.
}\label{Beta8hDL}
\end{figure}

The compensated spectrum $F(k) \cdot k^{5/2}$ is presented on Fig.\ref{Beta8hDL}. One can see plateau-like region responsible for $k^{-5/2}$ behavior, equivalent to $\omega^{-4}$ tail in Fig.\ref{LogLog8hDL}. This exact solution of Eq.(\ref{ConsHasEq}), known as KZ spectrum, was found by \cite{R4}. One should note that most of the energy flux into the system comes in the vicinity of the spectral peak, as shown on Fig.\ref{EnergyInputDurationLimited}, providing significant intertial interval for KZ spectrum.

\begin{figure}[ht]
\includegraphics[scale=0.7]{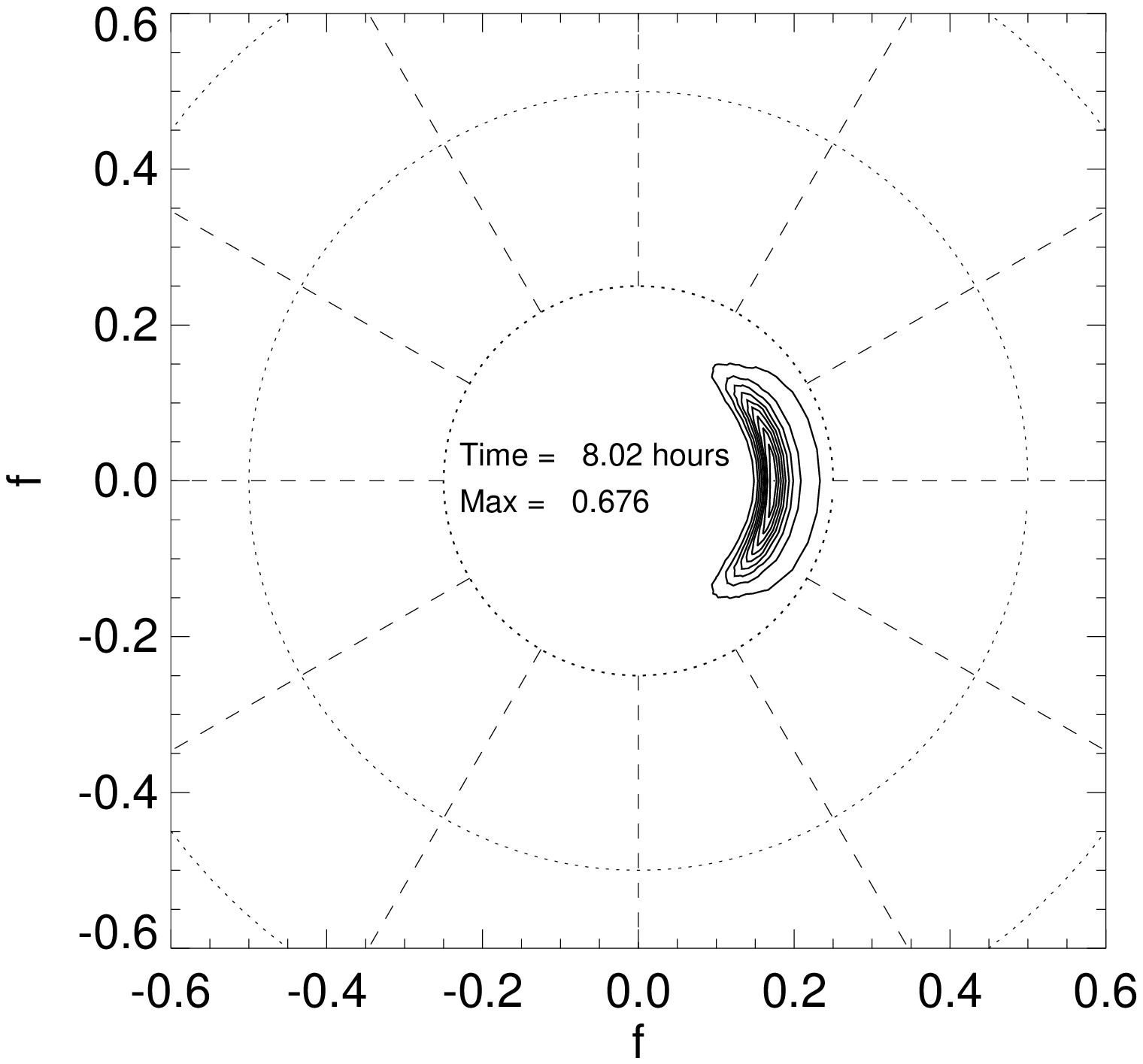}
\caption{
Angular spectra dependence for duration limited case.
}\label{Polar8hDL}
\end{figure}

The angular spectral distribution of energy, presented on Fig.\ref{Polar8hDL}, is consistent with the results of experimental observations that show a broadening of the angular spreading in both directions away from the spectral peak frequency.

Another important theoretical relationship, that can be derived from joint consideration of Eqs.  (\ref{ZKspec2}), (\ref{EnRel}) and (\ref{ZKspec1}):
\begin{eqnarray}
\label{TheorRegr}
1000\beta = 3 \frac{(u^2 c_p)^{1/3}}{g^{1/2}}
\end{eqnarray}
presents theoretical equivalent of the experimental regression.

\begin{figure}[ht]
\includegraphics[scale=0.7]{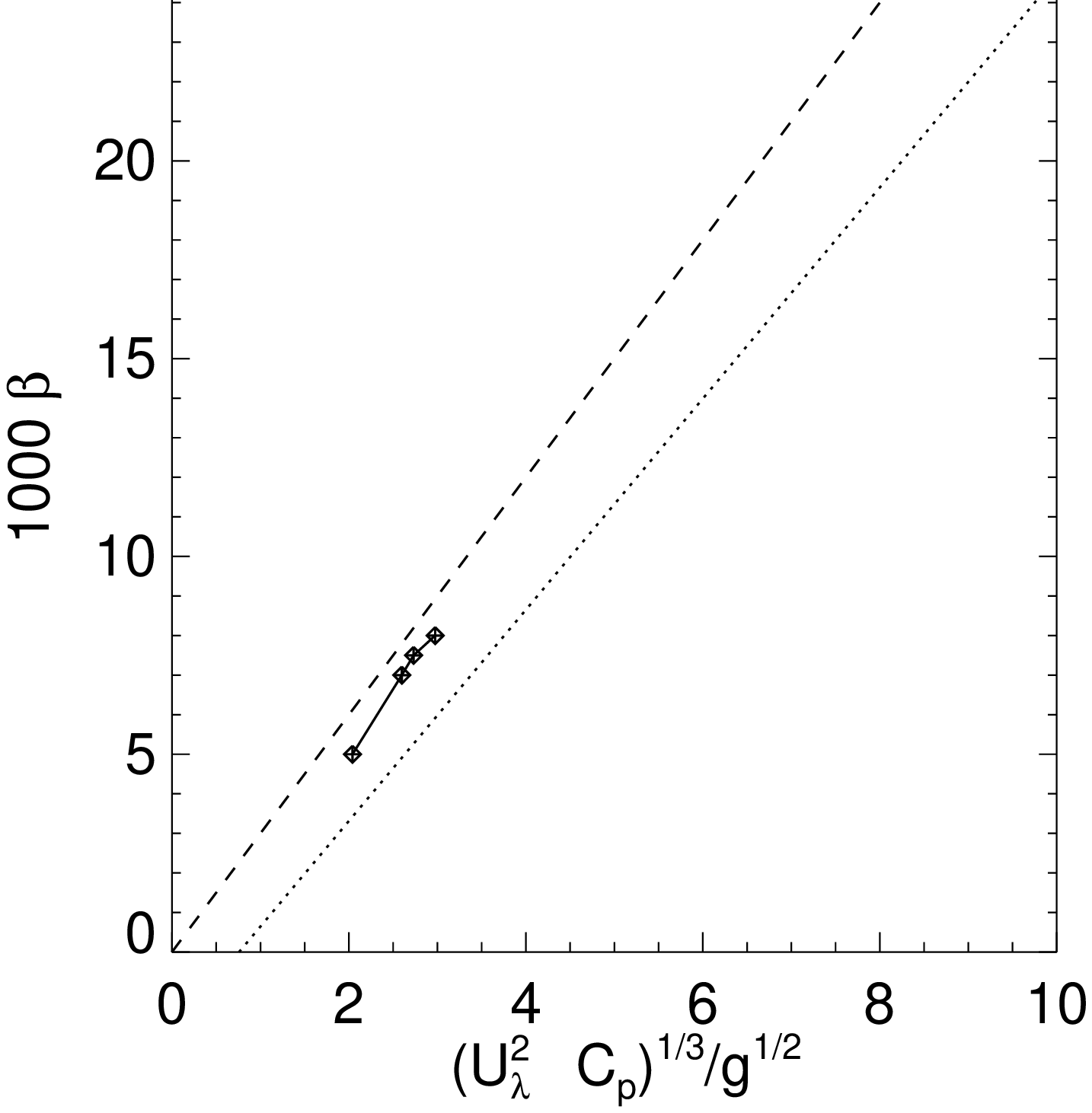}
\caption{Experimental, theoretical and numerical evidence of dependence of $1000 \beta$ on $\left(u_{\lambda}^2 c_p\right)^{1/3}/g^{1/2}$. Dashed line - theoretical prediction Eq.(\ref{TheorRegr}); dotted line - experimental regression line from \cite{R27,R20}. Diamonds - results of numerical calculations for duration limited case with wind speed $u_{10}=10$ m/sec.}
\label{BetaDL}
\end{figure}

To compare the performance of our numerical simulation  with the experimental analysis by \cite{R27}, presented in Fig.\ref{RegLine}, we show in Fig.\ref{BetaDL} plots of the compensated spectrum $\beta=F(k)\cdot k^{5/2}$ as a function of $(u_{\lambda}^2 C_p)^{1/3}/g^{1/2}$ for wind speeds $u_{10}=10.0$  m/sec, along with the regression line from \cite{R27}. The overall correspondence is quite good. 

\subsection{Limited fetch numerical simulations}

The limited fetch simulation was performed within the framework of the stationary version of the Eq.(\ref{HE}):

\begin{eqnarray}
\label{LFeq}
\frac{1}{2} \frac{g \cos \theta}{\omega } \frac{\partial \epsilon}{\partial x} = S_{nl}(\epsilon)+S_{wind}+S_{diss}
\end{eqnarray}

where $x$ is the coordinate orthogonal to the shore and $\theta$ is the angle between individual wavenumber $\vec{k}$ and the axis $\vec{x}$.

Stationarity in Eq.(\ref{LFeq}) is somewhat difficult for numerical simulation, since it contains the singularity in the form of $\cos \theta$ in front of $\frac{\partial \epsilon}{\partial x}$. We overcame this problem of division by zero through zeroing one half of the Fourier space of the system for the waves propagating toward the shore. Since the energy in such waves is small with respect to waves propagating in the offshore direction, such approximation is quite reasonable for our purposes.

Since the wind forcing index $s$ in the fetch-limited case is similar to that in the time domain, the numerical simulation of Eq.(\ref{LFeq}) has been performed for the same input functions as in the time doman case with the same low-level energy noise initial conditions in Fourier space.

\begin{figure}[t]
\includegraphics[scale=0.7]{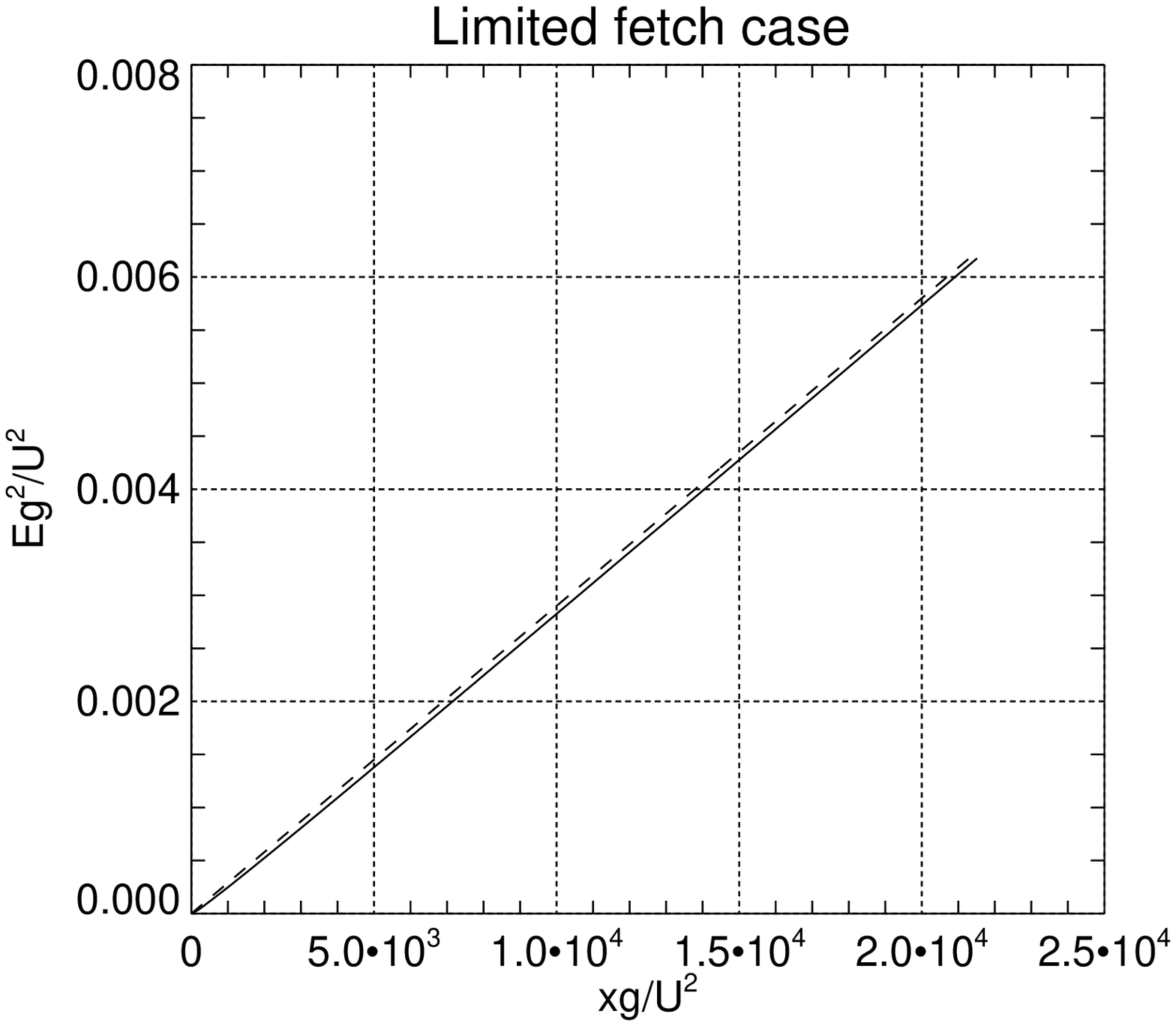}
\caption{
Dimensionless energy $Eg^2/U^4$ versus dimensionless fetch $xg/U^2$ for fetch limited case. Dashed fit $2.9 \cdot 10^{-7} xg/U^2$
}\label{EnergyDimFL}
\end{figure}

\begin{figure}[t]
\includegraphics[scale=0.7]{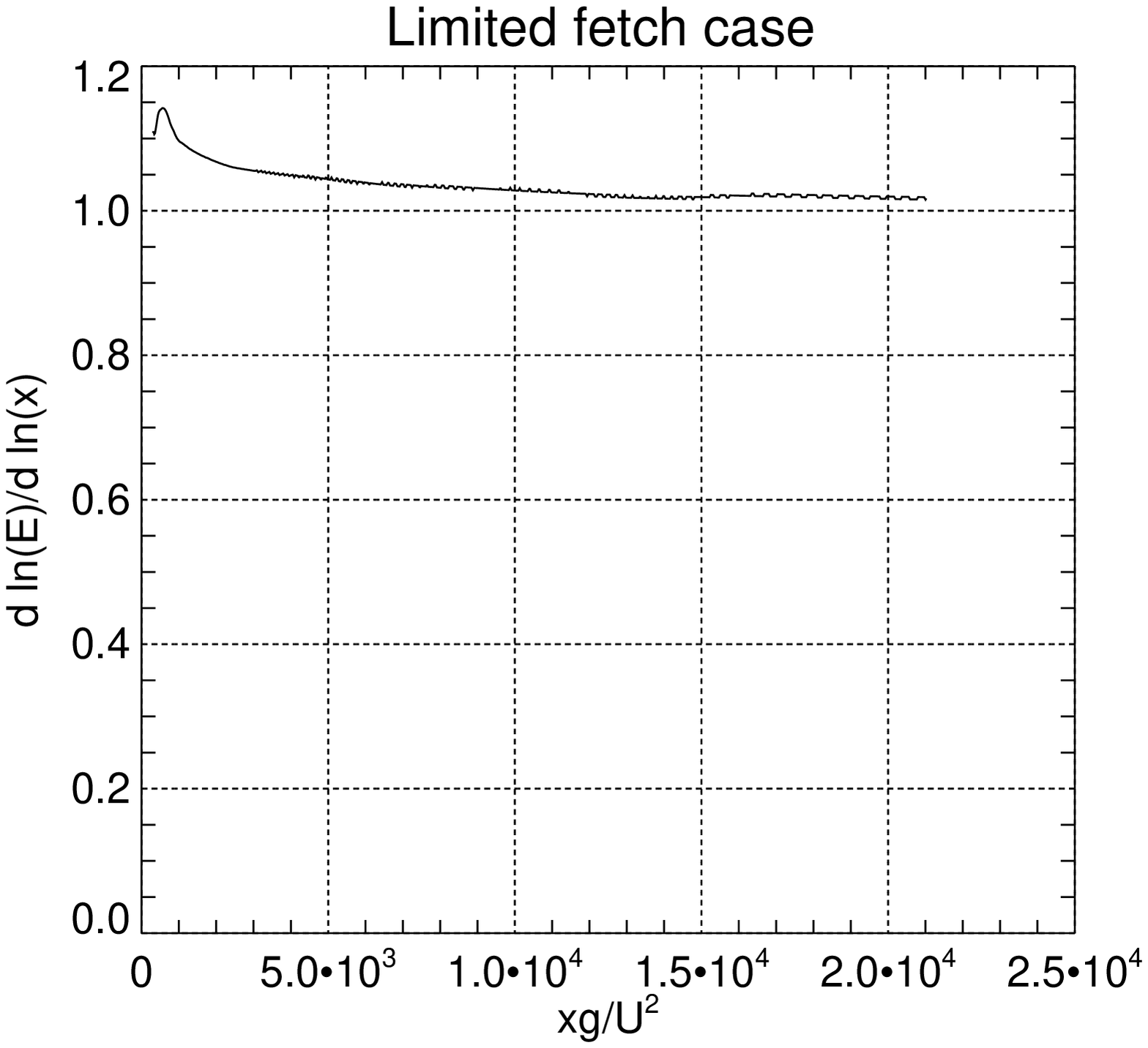}
\caption{
Local energy exponent power $p=\frac{d\ln{E} }{d\ln{x} }$ as a function of dimensionless fetch $xg/U^2$ for fetch limited case.
}\label{EnergyIndexFL}
\end{figure}

Fig.\ref{EnergyDimFL} shows total energy growth as a function of fetch coordinate close to self-similar prediction Eq.(\ref{EnFetchSelfSim}) for $p=1$, see Fig.\ref{EnergyIndexFL}.

\begin{figure}[ht]
\includegraphics[scale=0.7]{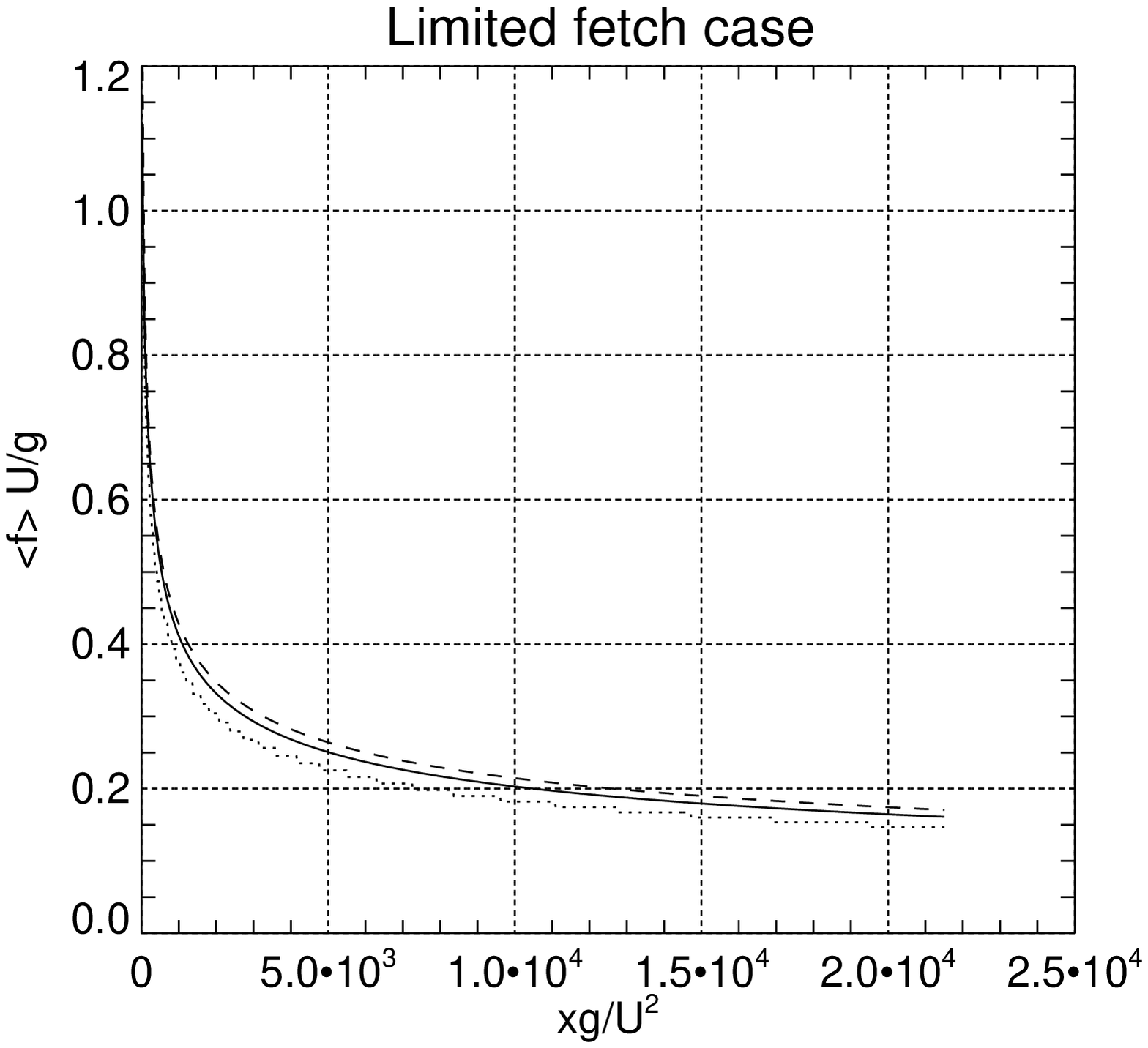}
\caption{Dimensionless mean frequency, as a function of dimensionless fetch (solid line), calculated as $<f>=\frac{1}{2\pi}\frac{\int\omega n d\omega d\theta}{\int n d\omega d\theta}$, where $n(\omega,\theta)=\frac{\varepsilon(\omega,\theta)}{\omega}$ is the wave action spectrum. The dotted line is the peak frequency $f_p=\frac{\omega_p}{2\pi}$, and the dashed is the theoretical fit $3.4\cdot \left(\frac{xg}{U^{2}} \right)^{-0.3}$. 
}\label{MeanFreqDimFL}
\end{figure}

\begin{figure}[ht]
\includegraphics[scale=0.7]{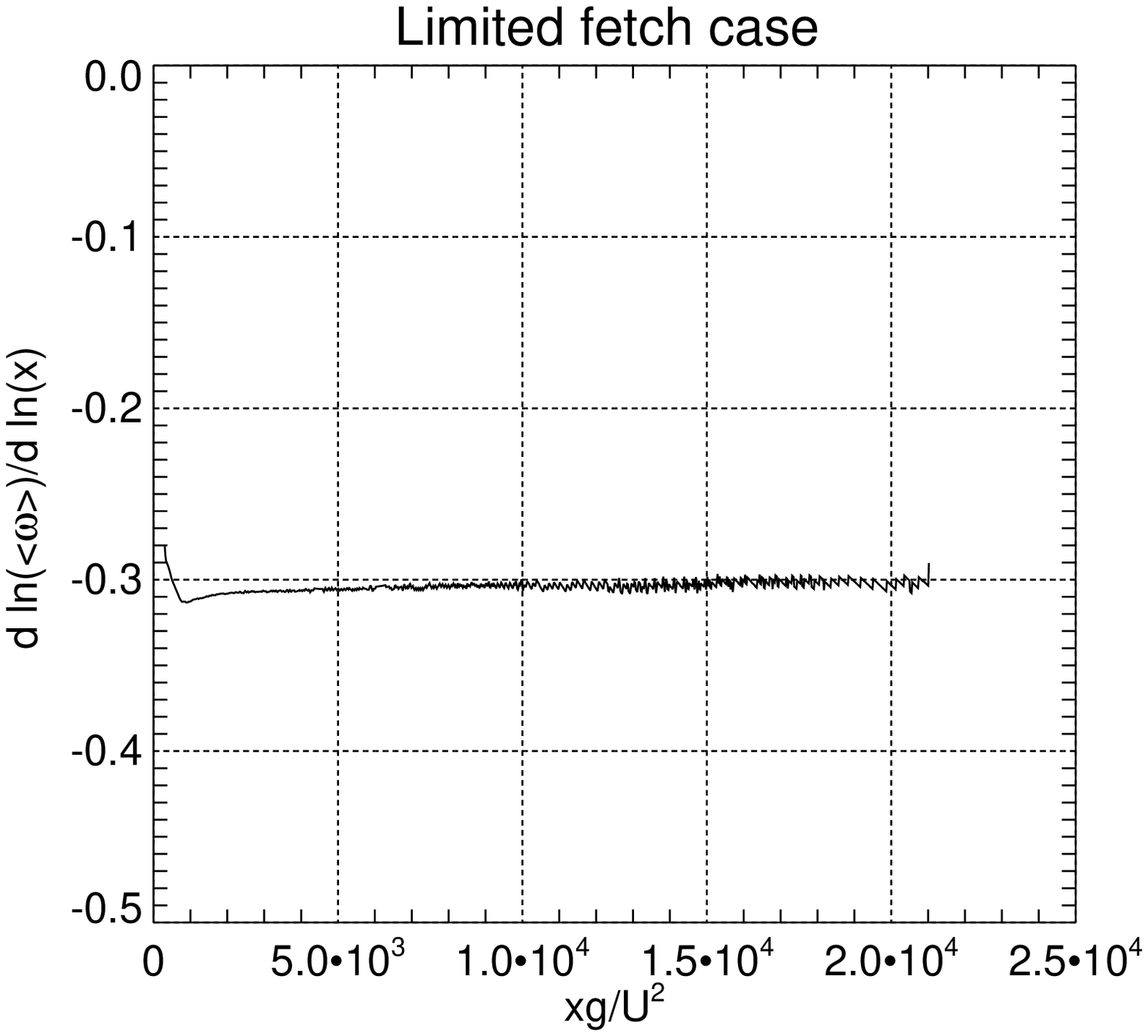}
\caption{
Local mean frequency exponent power $q=\frac{d\ln{<\omega>} }{d\ln{x} }$ as a function of dimensionless fetch $xg/U^2$ for fetch limited case.
}\label{MeanFreqIndexFL}
\end{figure}

Dependence of mean frequency on the fetch, shown on Fig.\ref{MeanFreqDimFL},  demonstrates  good correspondence with self-similar dependence Eq.(\ref{FrTimeSelfSim}), for $q = 3/10$, see Fig.\ref{MeanFreqIndexFL}.

\begin{figure}[ht]
\includegraphics[scale=0.7]{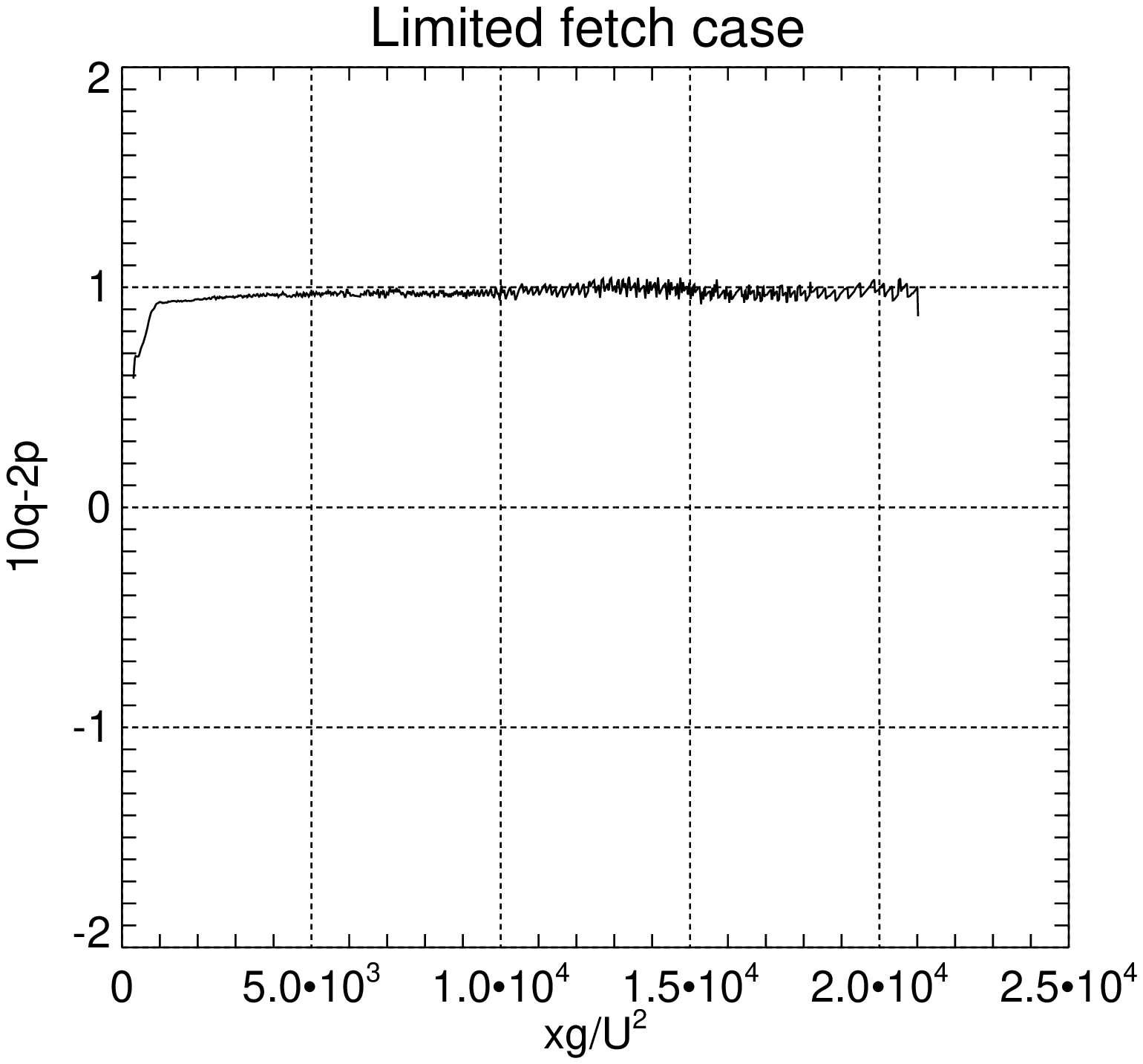}
\caption{
"Magic number" $10q-2p$ as a function of dimensionless fetch $xg/U^2$ for fetch limited case.
}\label{MagicNumberFL}
\end{figure}

The check of “magic number” $(10q-2p)$ is presented on Fig.\ref{MagicNumberFL}. It  agrees with the self-similar prediction of Eq.{\ref{MagicRelationFetch}}.

\begin{figure}[ht]
\includegraphics[scale=0.7]{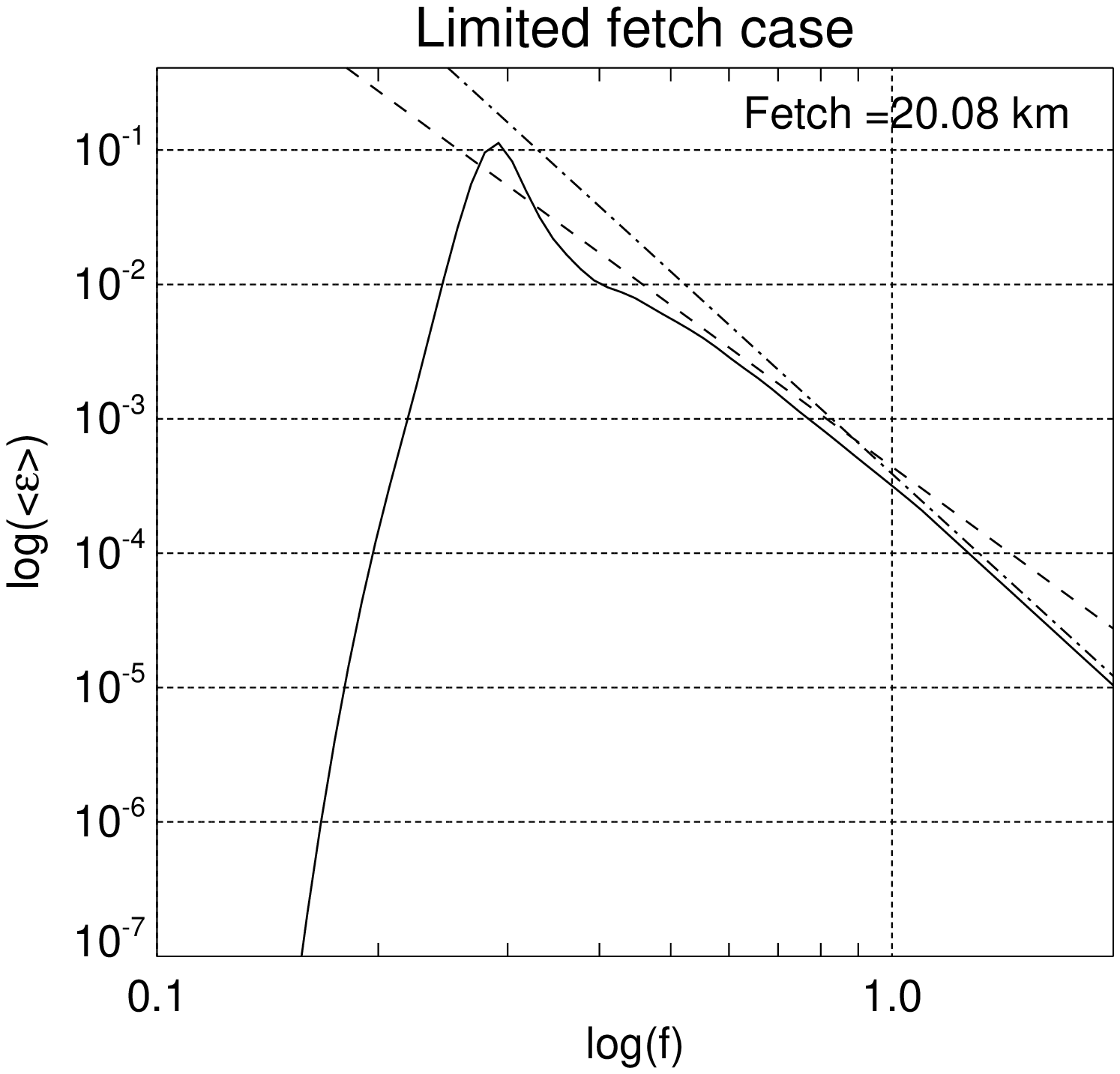}
\caption{
Decimal logarithm of the angle averaged spectrum as a function of the decimal logarithm of the frequency for fetch limited case.
}\label{LogLog20FL}
\end{figure}

Fig.\ref{LogLog20FL} presents directionally integrated energy spectrum, as the function of frequency, in logarithmic coordinates. As in the time domain case, one can see that it consists of the segments of:

\begin{itemize}
\item{} spectral maximum area
\item{} spectrum $~\omega^{-4}$
\item{} Phillips high frequency tail $\omega^{-5}$
\end{itemize}

\begin{figure}[ht]
\includegraphics[scale=0.7]{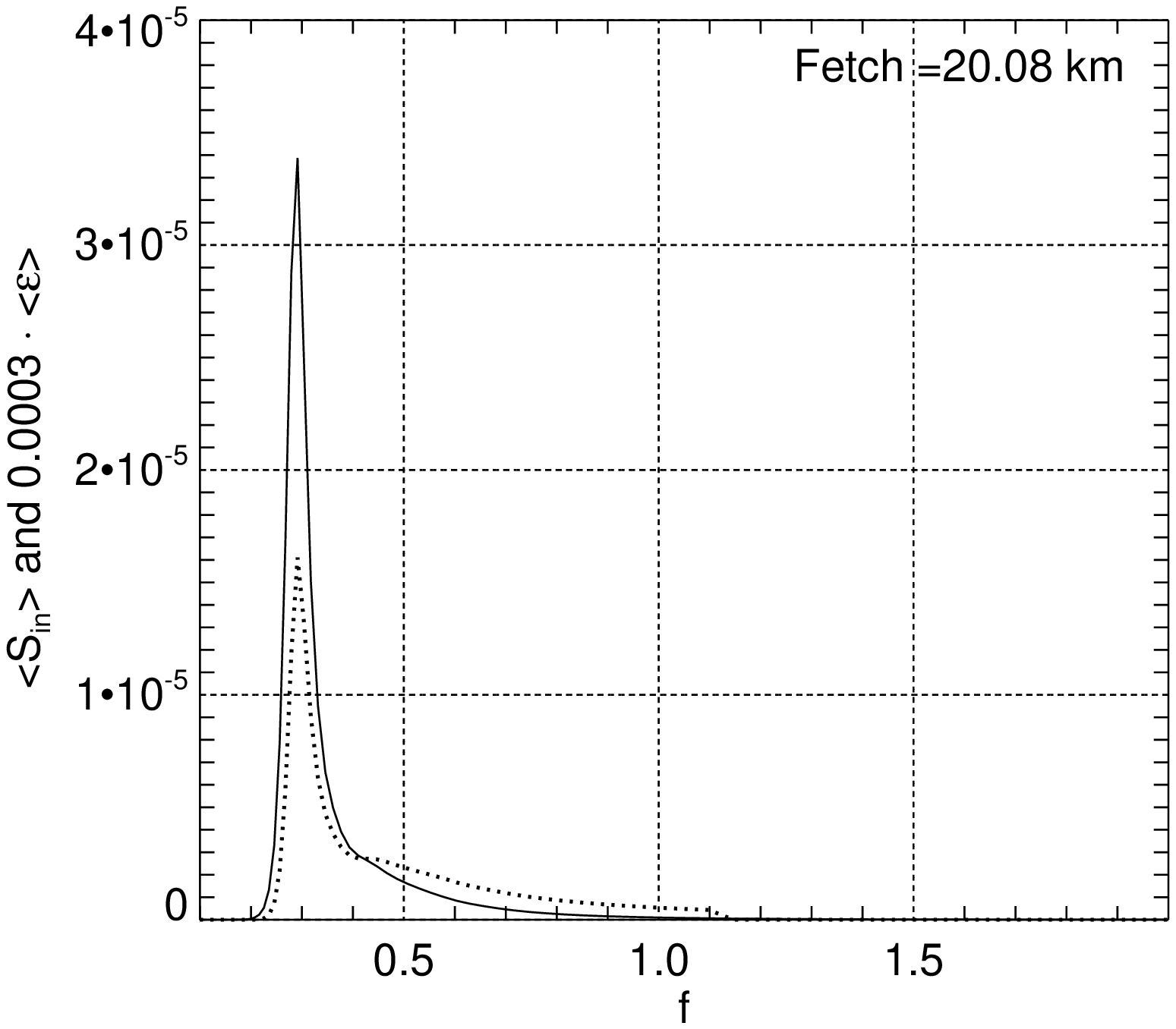}
\caption{
Typical, angle averaged, wind input function density $<S_{in}> = \frac{1}{2 \pi}\int \gamma(\omega,\theta) \varepsilon(\omega,\theta) d\theta$ and angle averaged spectrum $<\varepsilon> = \frac{1}{2\pi} \int \varepsilon(\omega,\theta) d\theta$ (solid line) as the functions of the frequency $f=\frac{\omega }{2\pi}$.
}\label{EnergyInputFetchLimited}
\end{figure}

\begin{figure}[ht]
\includegraphics[scale=0.7]{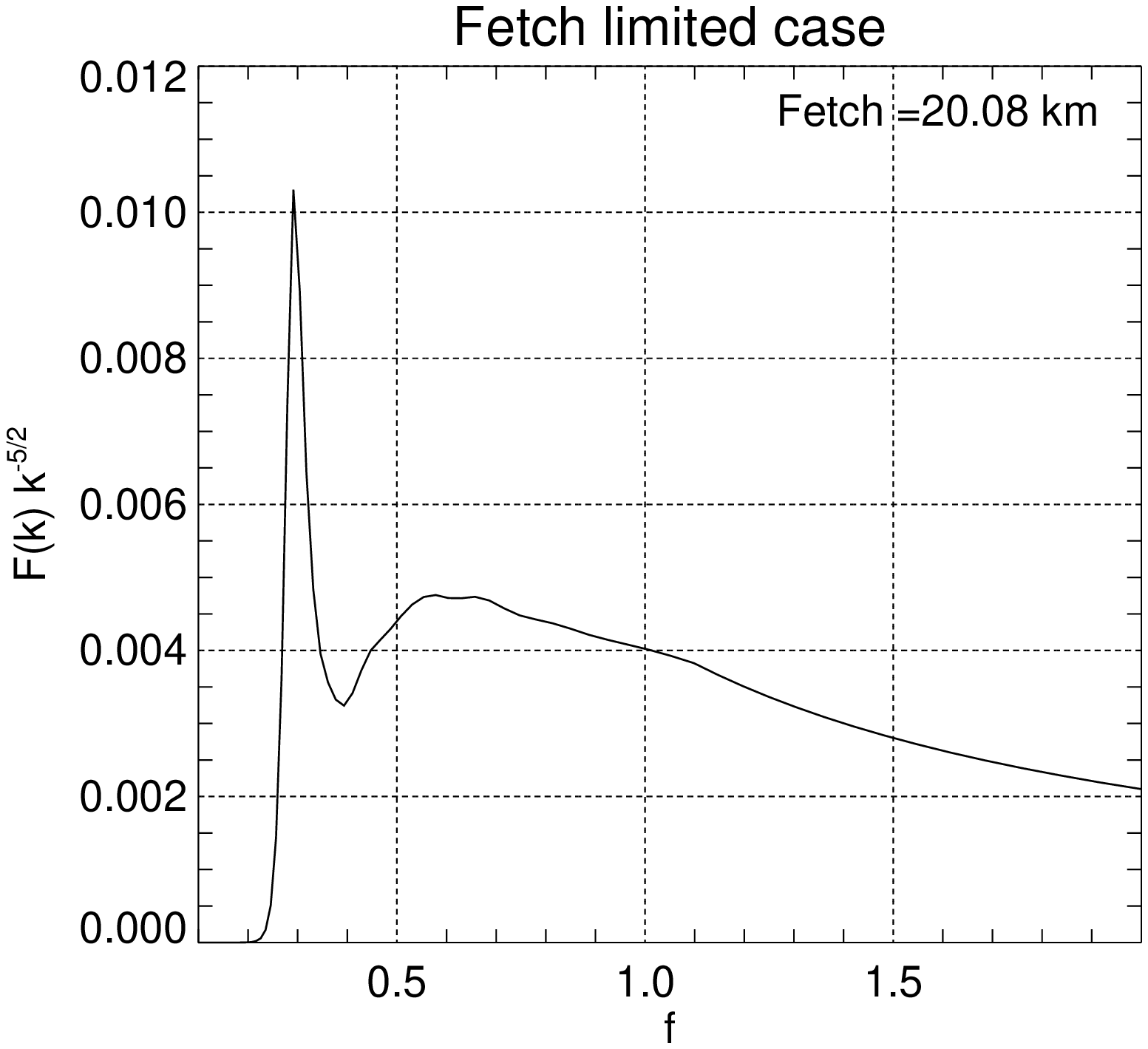}
\caption{
Compensated spectrum for fetch limited case as function of lnear frequency $f$.
}\label{Beta20FL}
\end{figure}

The compensated spectrum $F(k) \cdot k^{5/2}$ is presented on Fig.\ref{Beta20FL}. One can see plateu-like region responsible for $k^{-5/2}$ behavior, equivalent to $\omega^{-4}$ tail in Fig.\ref{LogLog20FL}. As in the time domain case, KZ solution \citep{R4} also holds for limited domain case, and most of the energy flux into the system comes in the vicinity of the spectral peak as well, as shown on Fig.\ref{EnergyInputFetchLimited}, providing significant intertial interval for KZ spectrum.

\begin{figure}[ht]
\includegraphics[scale=0.7]{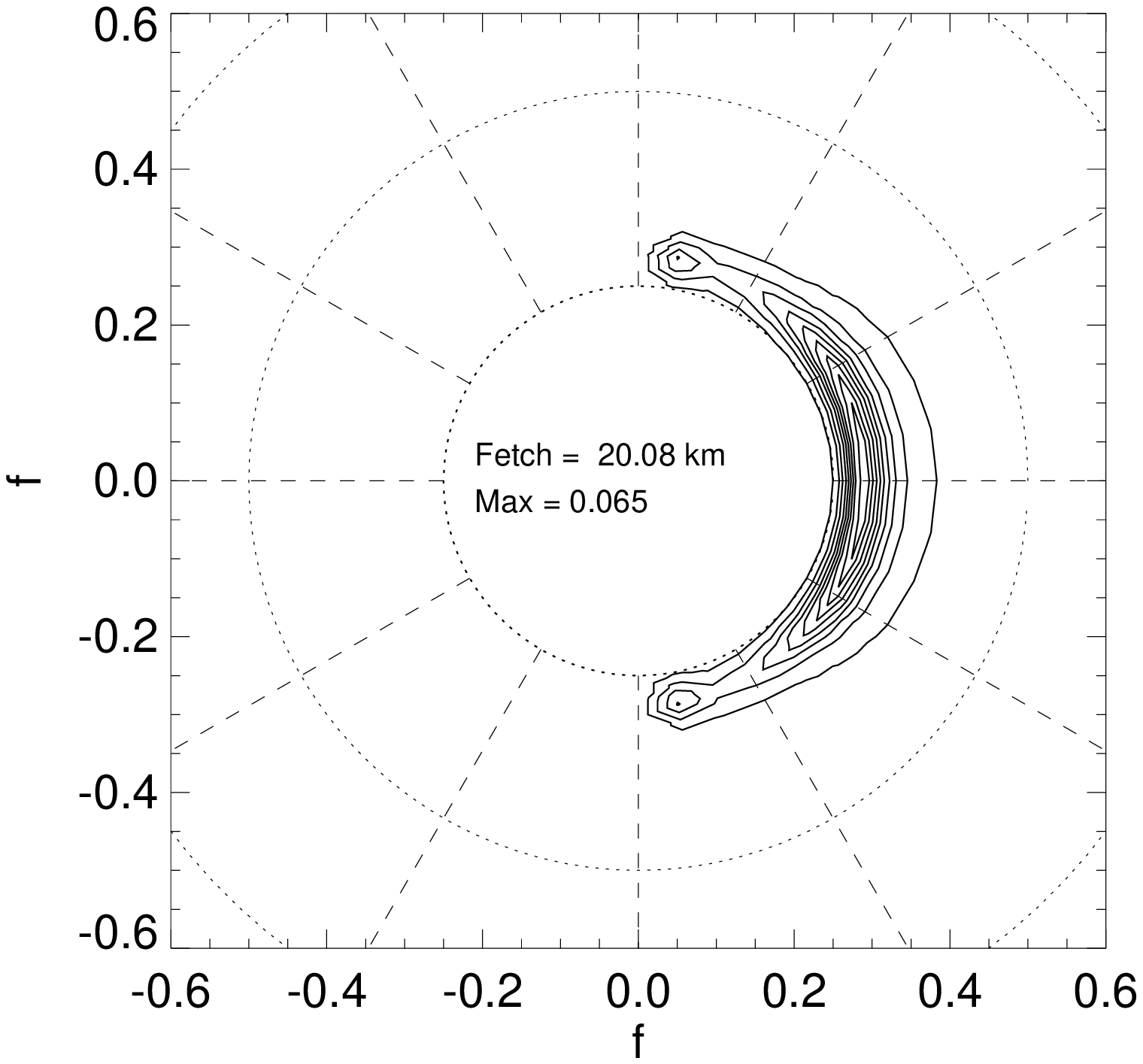}
\caption{
Angular spectra for fetch limited case.
}\label{Polar20FL}
\end{figure}

Angular spectrum distribution, consistent with the broadening of the spectrum in experimental observations is presented on Fig.\ref{Polar20FL}.

\begin{figure}[ht]
\includegraphics[scale=0.7]{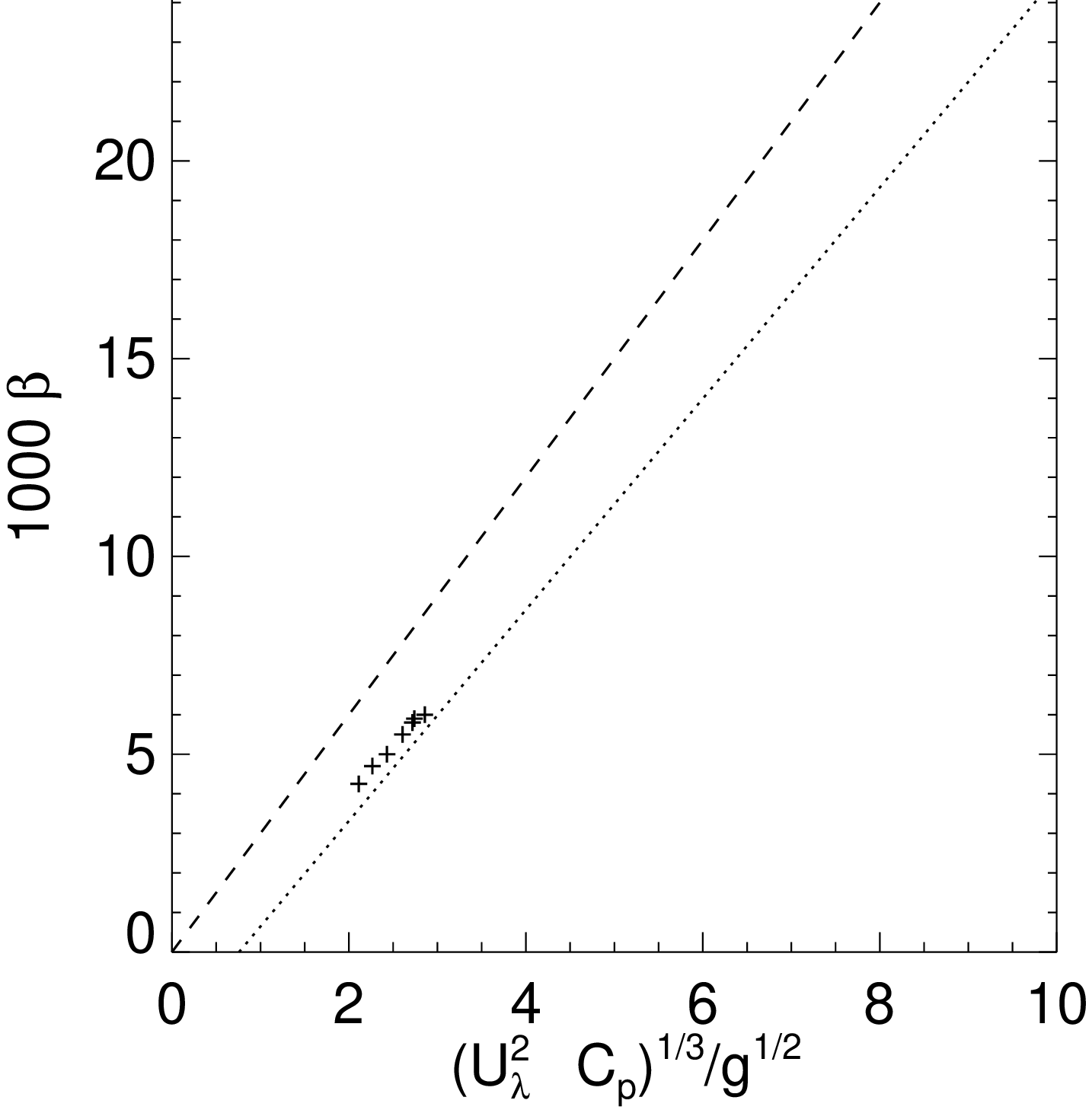}
\caption{Experimental and numerical evidence of dependence of $1000 \beta$ on $\left(u_{\lambda}^2 c_p\right)^{1/3}/g^{1/2}$.  Dashed line - theoretical prediction Eq.(\ref{TheorRegr}); dotted line - experimental regression line from \cite{R27,R20}. Crosses - results of numerical calculations for fetch limited case with wind speed $U=10$ m/sec.
}
\label{BetaFL}
\end{figure}

Fig.\ref{BetaFL} presents the plot of  $\beta=F(k)\cdot k^{5/2}$ as a function of $(u_{\lambda}^2 C_p)^{1/3}/g^{1/2}$ for wind speed $u_{10} = 10.0 $ m/sec, along with the regression line from \cite{R27} and theoretical prediction Eq.(\ref{TheorRegr}). The overall correspondence is quite good.

\section{Conclusion}

We  analysed the new ZRP form for wind input, proposed in \cite{R8}. Both numerical simulations for time domain and limited fetch cases, which used ZRP wind input term, XNL nonlinear term $S_{nl}$ and "implicit" high-frequency dissipation, shown good agreement with predicted self-similar properties of Hasselmann equation, experimentally obtained regression line \cite{R27,R20}and its theoretical prediction. The authors of the research hope that this new framework term will improve the quality of ocean wave prediction forecasts in operational models . 

\begin{acknowledgment} 

The research presented in sections 4a has been accomplished due to support
of the grant “Wave turbulence: the theory, mathematical modeling and
experiment” of the Russian Scientific Foundation No 14-22-00174. The research
presented in other chapters was supported by ONR grant N00014-10-1-0991.
The authors gratefully acknowledge the support of these foundations.

\end{acknowledgment}

% Create a bibliography directory and place your .bib file there.
% -REMOVE ALL DIRECTORY PATHS TO REFERENCE FILES BEFORE SUBMITTING TO THE AMS FOR PEER REVIEW
\ifthenelse{\boolean{dc}}
{}
{\clearpage}

\bibliographystyle{ametsoc}

\bibliography{references}

\begin{thebibliography}{24}
\providecommand{\natexlab}[1]{#1}
\providecommand{\url}[1]{\texttt{#1}}
\providecommand{\urlprefix}{URL }
\expandafter\ifx\csname urlstyle\endcsname\relax
  \providecommand{\doi}[1]{doi:\discretionary{}{}{}#1}\else
  \providecommand{\doi}{doi:\discretionary{}{}{}\begingroup
  \urlstyle{rm}\Url}\fi
\providecommand{\eprint}[2][]{\url{#2}}

\bibitem[{Badulin et~al.(2005)Badulin, Pushkarev, D.Resio, and Zakharov}]{R11}
Badulin, S.~I., A.~N. Pushkarev, D.Resio, and V.~E. Zakharov, 2005:
  Self-similarity of wind-driven sea. \textit{Nonlinear Proc. in Geophysics},
  \textbf{12}, 891 -- 945.

\bibitem[{Charnock(1955)}]{R26}
Charnock, H., 1955: Wind stress on a water surface. \textit{Q.J.R. Meteorol.
  Soc.}, \textbf{81}, 639 -- 640.

\bibitem[{Dyachenko et~al.(2015)Dyachenko, Kachulin, and Zakharov}]{R21}
Dyachenko, A.~I., D.~I. Kachulin, and V.~E. Zakharov, 2015: Evolution of
  one-dimensional wind-driven sea spectra. \textit{JETP Letters}, \textbf{102},
  577 -- 581.

\bibitem[{Golitsyn(2010)}]{R28}
Golitsyn, G., 2010: The energy cycle of wind waves on the sea surface.
  \textit{Izvestiya, Atmospheric and Oceanic Physics}, \textbf{46}, 6--13.

\bibitem[{Hasselmann(1962)}]{R1}
Hasselmann, K., 1962: On the non-linear energy transfer in a gravity-wave
  spectrum. {P}art 1. {G}eneral theory. \textit{Journal of Fluid Mechanics},
  \textbf{12}, 481 -- 500.

\bibitem[{Hasselmann(1963)}]{R2}
Hasselmann, K., 1963: On the non-linear energy transfer in a gravity wave
  spectrum. {P}art 2. {C}onservation theorems; wave-particle analogy;
  irrevesibility. \textit{Journal of Fluid Mechanics}, \textbf{15}, 273 -- 281.

\bibitem[{Irisov and Voronovich(2011)}]{R31}
Irisov, V. and A.~Voronovich, 2011: Numerical simulation of wave breaking.
  \textit{Journal of Physical Oceanography}, \textbf{41~(2)}, 346 -- 364.

\bibitem[{Longuet-Higgins(1980{\natexlab{a}})}]{R30}
Longuet-Higgins, M.~S., 1980{\natexlab{a}}: On the forming of sharp corners at
  a free surface. \textit{Proc.R.Soc.Lond. A}, \textbf{371}, 453 -- 478.

\bibitem[{Longuet-Higgins(1980{\natexlab{b}})}]{R29}
Longuet-Higgins, M.~S., 1980{\natexlab{b}}: A technique for time-dependent,
  free-surface flow. \textit{Proc.R.Soc.Lond. A}, \textbf{371}, 441 -- 451.

\bibitem[{Nordheim(1928)}]{R3}
Nordheim, L.~W., 1928: On the kinetic method in the new statistics and its
  application in the electron theory of conductivity. \textit{Proc. R. Soc.
  Lond. A}, \textbf{119}, 689 -- 698.

\bibitem[{Pushkarev and Zakharov(2016)}]{R7}
Pushkarev, A. and V.~Zakharov, 2016: Limited fetch revisited: comparison of
  wind input terms, in surface wave modeling. \textit{Ocean Modeling},
  \textbf{103}, 18 --– 37, \doi{10.1016/j.ocemod.2016.03.005}.

\bibitem[{Resio and Long(2007)}]{R20}
Resio, D.~T. and C.~E. Long, 2007: Wind wave spectral observations in
  {C}urrituck {S}ound, {N}orth {C}arolina. \textit{J. Geophys. Res.},
  \textbf{112}, C05\,001.

\bibitem[{Resio et~al.(2004)Resio, Long, and Vincent}]{R27}
Resio, D.~T., C.~E. Long, and C.~L. Vincent, 2004: Equilibrium-range constant
  in wind-generated wave spectra. \textit{J. Geophys. Res.}, \textbf{109},
  C01\,018.

\bibitem[{Resio and Perrie(1991)}]{R12}
Resio, D.~T. and W.~Perrie, 1991: A numerical study of nonlinear energy fluxes
  due to wave-wave interactions in a wave spectrum. {P}art {I}: Methodology and
  basic results. \textit{J. Fluid Mech.}, \textbf{223}, 603 -- 629.

\bibitem[{SWAN()}]{R10}
SWAN, 2015: {SWAN} team, \url{http://swanmodel.sourceforge.net/}.

\bibitem[{Tolman(2013)}]{R9}
Tolman, H.~L., 2013: User manual and system documentation of {WAVEWATCH III}.
  \textit{Environmental Modeling Center, Marine Modeling and Analysis Branch}.

\bibitem[{Zakharov(2005)}]{R17}
Zakharov, V.~E., 2005: Theoretical interpretation of fetch-limited wind-driven
  sea observations. \textit{NPG}, \textbf{13}, 1 -- 16.

\bibitem[{Zakharov(2010)}]{R14}
Zakharov, V.~E., 2010: Energy balances in a wind-driven sea. \textit{Physica
  Scripta}, \textbf{T142}, 014\,052.

\bibitem[{Zakharov and Badulin(2011)}]{R15}
Zakharov, V.~E. and S.~I. Badulin, 2011: On energy balance in wind-driven sea.
  \textit{Doklady Akademii Nauk}, \textbf{440}, 691 -- 695.

\bibitem[{Zakharov and Filonenko(1966)}]{R19}
Zakharov, V.~E. and N.~N. Filonenko, 1966: The energy spectrum for stochastic
  oscillation of a fluid's surface. \textit{Dokl.Akad.Nauk.}, \textbf{170},
  1992 -- 1995.

\bibitem[{Zakharov and Filonenko(1967)}]{R4}
Zakharov, V.~E. and N.~N. Filonenko, 1967: The energy spectrum for stochastic
  oscillations of a fluid surface. \textit{Sov. Phys. Docl.}, \textbf{11}, 881
  -- 884.

\bibitem[{Zakharov et~al.(2009)Zakharov, Korotkevich, and Prokofiev}]{R22}
Zakharov, V.~E., A.~O. Korotkevich, and A.~O. Prokofiev, 2009: On dissipation
  function of ocean waves due to whitecapping. \textit{American Institute of
  Physics Conference Series}, T.~E. Simos, G.Psihoyios, and C.~Tsitouras, Eds.,
  Vol. 1168, 1229 -- 1237.

\bibitem[{Zakharov et~al.(1992)Zakharov, L'vov, and Falkovich}]{R5}
Zakharov, V.~E., V.~S. L'vov, and G.~Falkovich, 1992: \textit{Kolmogorov
  Spectra of Turbulence I: Wave Turbulence}. Springer-Verlag.

\bibitem[{Zakharov et~al.(2012)Zakharov, Resio, and Pushkarev}]{R8}
Zakharov, V.~E., D.~Resio, and A.~Pushkarev, 2012: New wind input term
  consistent with experimental, theoretical and numerical considerations.
  http://arxiv.org/abs/1212.1069/.

\end{thebibliography}

\end{document}